\documentclass{article}

\usepackage{arxiv}

\usepackage[utf8]{inputenc} 
\usepackage[T1]{fontenc}    
\usepackage{hyperref}       
\usepackage{url}            
\usepackage{booktabs}       
\usepackage{amsfonts}       
\usepackage{nicefrac}       
\usepackage{microtype}      
\usepackage{lipsum}
\usepackage{graphicx}
\usepackage{xcolor}
\usepackage{amssymb}

\providecommand{\bo}{\mathbf}

\providecommand{\cov}{\mathrm{Cov}}

\providecommand{\diag}{\mathrm{diag}}
\providecommand{\off}{\mathrm{off}}

\newcommand{\indep}{\bot\!\!\,\!\!\bot}

\title{Stationary subspace analysis based on second-order statistics}

\author{
 Lea Flumian \\
  Private\\
   \And
 Markus Matilainen \\
  University of Turku\\
  \AA bo Akademi University\\
  \And
 Klaus Nordhausen \\
  University of Jyv\"askyl\"a\\
   \And
 Sara Taskinen \\
  University of Jyv\"askyl\"a\\
}

\begin{document}
\maketitle

\begin{abstract}
In stationary subspace analysis (SSA) one assumes that the observable $p$-variate time series is a linear mixture of a $k$-variate nonstationary time series and a $(p-k)$-variate stationary time series. The aim is then to estimate the unmixing matrix which transforms the observed multivariate time series onto stationary and nonstationary components. In the classical approach multivariate data are projected onto stationary and nonstationary subspaces by minimizing a Kullback–Leibler divergence between Gaussian distributions, and the method only detects nonstationarities in the first two moments. In this paper we consider SSA in a more general multivariate time series setting and propose SSA methods which are able to detect nonstationarities in mean, variance and autocorrelation, or in all of them. Simulation studies illustrate the performances of proposed methods, and it is shown that especially the method that detects all three types of nonstationarities performs well in various time series settings. The paper is concluded with an illustrative example.
\end{abstract}


\section{Introduction}
\label{S:1}

Multivariate time series data are observed in many application areas and are often very challenging to model. To ease the analyses, it is often assumed that the observed time series can be decomposed into latent components with different exploitable properties. One common approach is to apply one of the blind source separation (BSS) methods to observed data in order to estimate the latent components as a pre-processing tool. In a special case of BSS, that is, in second-order source separation (SOS), it is assumed that the latent components are uncorrelated second-order stationary time series, and in independent component time series model, the assumption on uncorrelatedness is replaced with the assumption on independence (see {\it e.g.} \cite{Miettinenetal:2016}, and references therein). In nonstationary source separation (NSS), as defined in \cite{ChoiCichocki:2000}, the variances of uncorrelated sources are allowed to change over time. For a recent review of these and other variants of BSS, see \cite{ComonJutten:2010,NordhausenOja2018,PanMatilainenTaskinenNordhausen2021}.  

In stationary subspace analysis (SSA), the underlying model states that the observable $p$-variate time series is a linear mixture of a $k$-variate nonstationary time series and  a $(p-k)$-variate stationary time series. The aim is then to factorize the multivariate time series onto stationary and nonstationary components. Such a factorization is useful in several real world applications, {\it e.g.}, in biomedical signal processing, speech recognition, image analysis and econometrics, where either the stationary or the nonstationary components are of interest. SSA was introduced in~\cite{BunauMeineckeKiralyMuller:2009}, where the matrix that separates the subspaces of stationary and nonstationary components was found by dividing the observed multivariate time series into $K$ segments and minimizing a Kullback-Leibler (KL) divergence between Gaussian distributions thus measuring differences in means and covariances across segments. We denote such method as KL-SSA. Notice that in KL-SSA the time series are considered as stationary if the first two moments are time-invariant. The stationarity with respect to the autocorrelations is not taken into account. \cite{BunauMeineckeKiralyMuller:2009} discussed the theoretical properties of the proposed method and proposed a sequential likelihood ratio test for testing the  dimension of the stationary subspace. The method was applied to the EEG data analysis in \cite{BunauMeineckeKiralyMuller:2009,Bunauetal:2010} and to computer vision in \cite{Meinecketeal:2009}, and extended to change-point detection in \cite{BlytheBunauMeineckeMuller:2012,Blytheetal:2013}, where the nonstationary components were of the main interest. In \cite{Horevetal:2016,Kaltenstadleretal:2018} several alternatives for the Kullback-Leibler divergence in SSA were given.

In this paper we propose a novel SSA algorithm that detects nonstationarities in the first two moments as well as in autocorrelations. The method can be seen as an extension of analytic SSA (ASSA), which is an SSA method suggested in \cite{Haraetal:2010}, to a more general multivariate time series setting. Notice that \cite{Sundarajanetal:2017} considered an alternative extension via a frequency domain approach. As another contribution we make a connection between SSA and supervised dimension reduction (SDR) methods. The paper is organized as follows. In Section~\ref{sec:SSA} we first review the model assumptions underlying SSA and discuss the two classical methods for performing SSA, that is, KL-SSA and ASSA. We also show how SSA can be seen as a supervised dimension reduction method and propose and compare methods for detecting different type of nonstationarities. Section~\ref{sec:practical} discusses practical issues and identifiability of stationary and nonstationary components. Section~\ref{sec:simulations} presents several simulation studies for comparing performances of different methods under various scenarios and an example. The paper is concluded with some discussion in Section~\ref{sec:discussion}.  

\section{Stationary subspace analysis}
\label{sec:SSA}

In this section we review the model underlying SSA, show some connections between SSA and supervised dimension reductions methods, and suggest methods to detect changes in mean, variance, correlation structure or in all of them. Tools to identify the type of nonstationarity the components exhibit are also given.

\subsection{Model formulation and notations}


Let $\bo{x_t}$ be an observable  $p$-variate  nonstationary time series which can be decomposed into a stationary part and a nonstationary part according to the SSA model as defined in \cite{BunauMeineckeKiralyMuller:2009}
\begin{equation}
\label{eq::SSAmodel}
\bo x_t = \bo A \bo z_t = [\bo A_{\bo s}\  \bo A_{\bo n} ] \left(
                                                           \begin{array}{c}
                                                             \bo s_t \\
                                                             \bo n_t \\
                                                           \end{array}
                                                         \right),
\end{equation}
where a $p$-variate latent time series $\bo z_t$ consists of a $k$-variate nonstationary time series $\bo n_t$ and a $(p-k)$-variate stationary time series $\bo s_t$. The two components are mixed using a full-rank $p \times p$ matrix $\bo A$. Here matrices $\bo A_{\bo s}$ and $\bo A_{\bo n}$ are $p\times (p-k)$ and $p\times k$ matrices, respectively. The aim of SSA is to estimate the unmixing matrix $\bo W=\bo A^{-1}$ so that $\bo W\bo x_t$ is partitioned into stationary and nonstationary time series. Depending on the assumptions on $\bo s_t$ and $\bo n_t$, different ways to estimate $\bo W$ are suggested in the literature \cite{BunauMeineckeKiralyMuller:2009,Haraetal:2010}. In this paper we make the following assumptions on the latent time series:
\begin{itemize}
  \item[(A1)] $E(\bo s_t)= \bo 0$, $\cov(\bo s_t) = \bo I_{p-k}$ and $\cov(\bo s_t,\bo s_{t+\tau})<\infty$,
  \item[(A2)] $E(\bo n_t)<\infty$ and $\cov(\bo n_t) = \bo D_t$ where $\bo D_t$ is a diagonal matrix with positive diagonal elements, 
\item[(A3)] $\cov(\bo s_t, \bo n_{t+\tau}) = \bo 0$  for all $\tau \geq 0$.
\end{itemize}
We thus assume that the $(p-k)$ time series in $\bo s_t$ are second-order stationary with finite autocovariances. The first assumption fixes the location and covariance matrix of the stationary part for convenience. The second assumption states that the $k$ nonstationary components in $\bo n_t$ have finite first moments and are uncorrelated, and the third assumption states that the nonstationary components are uncorrelated with the stationary components. 

Despite the assumptions the model is not well-defined as there are many divisions of $\bo x_t$ such that the assumptions hold. Here we are interested in the one with the minimal value of $k$. The stationary components are only specified up to a rotation by an orthogonal matrix, whereas the nonstationary components can be marginally rescaled, shifted and also rotated. Therefore for convenience of presentation we make the following additional assumption on $\bo n_t$:
\begin{itemize}
  \item[(A4)] $\sum_{t\in T} E(\bo n_t)= \bo 0$ and $\sum_{t \in T} \cov(\bo n_t) = \sum_{t \in T} \bo D_t = \bo I_k$.
\end{itemize}
Hence for the observed time span $T$ the location and scale are considered as fixed. 

The common idea in \cite{BunauMeineckeKiralyMuller:2009,Haraetal:2010} was to divide the observed time series into $K$ segments, then compute for each interval the first and the second order statistics and to measure their nonconstancy across segments. To be more exact, assume that $\bo x_t$ is observed at time points $1,\ldots,T$, and let $T_1,\ldots,T_K$ denote $K$ disjoint subsets of $T$, {\it e.g.}, non-overlapping intervals. Let then
\[
\bo m_{T_i}(\bo x_t) = \frac{1}{|T_i|} \sum_{t\in T_i} \bo x_t
\]
and 
\[
\bo S_{\tau,T_i}(\bo x_t)= \frac{1}{|T_i|-\tau} \sum_{t\in T_i} (\bo x_t - \bo m_{T_i}(\bo x_t)) (\bo x_{t+\tau} - \bo m_{T_i}(\bo x_t))^\top
\]
denote the sample mean and the sample (auto)covariance matrix computed using the data from interval $T_i$, respectively. In \cite{BunauMeineckeKiralyMuller:2009,Haraetal:2010} the unmixing matrix revealing the stationary sources was chosen so that the means and covariance matrices vary the least across all intervals. It was also shown that using the standardised time series defined as
\begin{equation}
\label{stand}
\bo y_t = \bo S_{0,T}(\bo x_t)^{-1/2}(\bo x_t - \bo m_{T}(\bo x_t)),
\end{equation}
the search can be restricted to orthogonal matrices only. In \cite{BunauMeineckeKiralyMuller:2009} KL divergence between the Gaussian distributions on each intervals and the standard normal distribution was minimized. Further, in \cite{Haraetal:2010} a second-order Taylor approximation was applied to the objective function of KL-SSA and the SSA problem was shown to reduce to a simple generalized eigenvalue problem. We review the resulting analytic SSA (ASSA) method in Section~\ref{sec:comb}. Notice that in both approaches the focus was on detecting stationarity deviations in mean and in variance but not in autocorrelation.

\subsection{Connection to supervised dimension reduction}
\label{sec:supervised}

Before going into specific SSA methods we point out a connection between SSA and supervised dimension reduction (SDR). In supervised dimension reduction, we have a response $z$ and a $p$-variate predictor vector $\bo x$ and the goal is to find a $k \times p$ matrix $\bo W$ such that $\bo W^\top \bo x$ carries all relevant information about $z$, that is, $\bo x \indep z\,|\,\bo W^\top \bo x$ with the smallest possible value of $k \ll p$. See \cite{MaZhu2013,Li:2018} for some general overviews. Two popular methods in this context are sliced inverse regression (SIR) \cite{Li:1991}  and sliced average variance estimation (SAVE) \cite{Cook:2000}. For a sample $(z_j, \bo x_j)$, $j=1,\ldots,n$, both methods start by whitening $\bo x_j$, that is, by computing $\bo y_j = \bo S_{0,n}^{-1/2}(\bo x_j-\bo m_n(\bo x_j))$ and then group the whitened observations into $K$ so-called slices $H_1,\ldots, H_K$ according to their response values $z_i$. By a slight abuse of the notation from above one is then interested in the slice means $\bo m_i(\bo y)=\frac{1}{|H_i|} \sum_{i\in H_i} \bo y_i$ and slice covariance matrices $\bo S_i(\bo y) = \frac{1}{|H_i|} \sum_{i \in H_i} (\bo y_i-\bo m_i)(\bo y_i-\bo m_i)^\top$, where $i=1,\ldots, K$. SIR estimate is then obtained using the covariance of the slice means and SAVE estimate is based on the average of the differences between the slice covariance matrices and the covariance matrix of all $\bo y_i$ which is due to the whitening $\bo I_p$.

\cite{CookCritchley2000} show that SAVE is more comprehensive than SIR in detecting the subspace of interest and give situations where SIR fails. The increased flexibility of SAVE comes however at the cost of requiring more data and being more sensitive to the number of slices \cite{CookCritchley2000}. Thus, while theoretically superior, in practice SIR is still preferred. Also combinations of SIR and SAVE are regularly investigated, see for example \cite{YeWeiss2003,ZhuOhtakiLi2007,ShakerPrendergast2011} for more details.

If the response $z_i$ is categorical, slicing is usually done by the unique values of $z_i$ while for numeric values of $z_i$ the slices are usually chosen so that they contain approximately the same number of observations. In this case the number of slices has to be chosen and it is a trade-off between trying to have many slices to find the directions and to have enough data points in the slice in order to estimate the slice statistic sufficiently well. SIR is considered quite robust with regard to the selection of number of slices while SAVE is considered quite sensitive. For example for SIR the number of slices should exceed the dimension of the subspace to be detected. 

In the context of SSA one can now think of the intervals as the response, where $z$ gets a distinct value in each different interval. Then naturally only the nonstationary components carry information about the ``response''.  In the following sections we consider SSA methods that correspond to sliced inverse regression (SIR) \cite{Li:1991} and sliced average variance estimation (SAVE) \cite{Cook:2000}, respectively. Section~\ref{sec:cor} proposes a method for detecting
stationarity deviations in autocorrelation as these cannot be detected by SIR and SAVE type approaches. Further, inspired by the success of hybrid methods in SDR we also consider a combined approach in Section~\ref{sec:comb}.

Note that SDR methods like SIR and SAVE and combinations of these have also been extended to the time series settings, see \cite{MatilainenCrouxNordhausenOja2017,MatilainenCrouxNordhausenOja2019}. However, these methods again focus on regression modeling and not for detecting nonstationary subspaces.

\subsection{Nonstationarity in mean}
\label{sec:mean}

Consider first a SSA method that aims at detecting stationary deviations in mean. Examples of such nonstationary components include trend, seasonal and cyclic components. Assume now that $\bo x_t$ is generated from the SSA model~(\ref{eq::SSAmodel}) and write $\bo y_t$ for the time series standardised using~(\ref{stand}). Let then 
\begin{equation}
\label{eq:mean}
\bo M_m = \sum_{i=1}^{K} \frac{|T_i|}{T}\bo m_{T_i}(\bo y_t) \bo m_{T_i}(\bo y_t)^\top
 \end{equation}
be the covariance between the means of the different intervals weighted by the interval length. Notice that with seasonal and cyclic components, the intervals have to be chosen so that they do not correspond to the cycles. This is also illustrated in Section~\ref{sec:CompEx}. 

It is easy to see that the eigenvalues of $\bo M_m$ that correspond to the components that are stationary in mean should be zero. Further, the components with non-zero eigenvalues correspond to components that are nonstationary in mean. Hence write the eigenvalue decomposition of $\bo M_m$ as
\[
\bo M_m = \bo U_m \bo D_m \bo U_m^\top,
\]
where $\bo D_m$ is a $p\times p$ diagonal matrix with the eigenvalues of $\bo M_m$ as diagonal values, and $\bo U_m = (\bo U_{m,s}^\top, \bo U_{m,n}^\top)^\top$ includes the eigenvectors of $\bo M_m$ arranged so that $\bo U_{m,n}$ is the $k_m \times p$ matrix containing the eigenvectors belonging to the non-zero eigenvalues as rows and $(p-k_m) \times p$ matrix $\bo U_{m,s}$ includes the remaining ones. 
The columns of resulting unmixing matrices $\bo W_{m,n} = \bo U_{m,n} \bo S_{0,T}(\bo x_t)^{-1/2}$ and $\bo W_{m,s} = \bo U_{m,s} \bo S_{0,T}(\bo x_t)^{-1/2}$ then generate the nonstationary and stationary subspaces, respectively. Naturally $k_m \leq k$ with equality only if for all nonstationary components the means differ at least between some of the chosen intervals. As this method is based on interval means it corresponds basically to SIR and we denote this method accordingly SSAsir.

\subsection{Nonstationarity in variance}
\label{sec:var}

Let us then consider a SAVE type method for detecting stationary deviations in variance. Consider again standardised time series $\bo y_t$ and let now
\begin{equation}
\label{eq:var}
\bo M_v = \sum_{i=1}^{K} \frac{|T_i|}{T} (\bo I_p- \bo S_{0,T_i}(\bo y_t))^2,
\end{equation}
where $\bo A^2=\bo A\bo A^\top$, measure the deviation of the covariance computed on intervals $T_1,\dots,T_K$ from the global covariance $\bo I_p$. 

Again the eigenvalues of $\bo M_v$ that correspond to the components that are nonstationary in variance should be non-zero. Let thus the eigenvalue decomposition be
\[
\bo M_v = \bo U_v \bo D_v \bo U_v^\top,
\]
where $\bo D_v$ is a $p\times p$ diagonal matrix with the eigenvalues of $\bo M_v$ as diagonal values, $\bo U_v = (\bo U^\top_{v,s}, \bo U^\top_{v,n})^\top$ are the eigenvectors of $\bo M_v$ arranged so that $\bo U_{v,n}$ is the $k_v \times p$ matrix containing the eigenvectors belonging to the non-zero eigenvalues as rows and and $\bo U_{v,s}$ is the $(p-k_v) \times p$ matrix containing the remaining eigenvectors as rows. Again $k_v \leq k$. The transforming matrices to the two subspaces are accordingly $\bo W_{v,n} = \bo U_{v,n} \bo S_{0,T}(\bo x_t)^{-1/2}$ and $\bo W_{v,s} = \bo U_{v,s} \bo S_{0,T}(\bo x_t)^{-1/2}$. As illustrated in Section~\ref{sec:CompEx}, although this approach is designed to detect components with nonstationary variances, it may also detect components which are nonstationary in mean as if the mean changes the variances in intervals might also change. We will refer to this method in the following as SSAsave.

\subsection{Nonstationarity in autocorrelation}
\label{sec:cor}

The SIR and SAVE type methods are able to detect components which are nonstationary with respect to the first and the second moments. To detect nonstationarities in autocorrelation we need a statistic to measure the nonconstancy in autocorrelation across segments. Let such statistic be defined (for standardised time series) as  
\begin{equation}
\label{eq:cor}
\bo M_\tau = \sum_{i=1}^{K} \frac{|T_i|}{T} (\bo S_{\tau,T}(\bo y_t)- \bo S_{\tau,T_i}(\bo y_t))^2,
\end{equation}
where $\tau \geq 1$. The matrix $\bo M_\tau$ thus measures the deviation of the autocovariance matrix computed on intervals $T_1,\dots, T_K$ from the global autocovariance matrix $\bo S_{\tau,T}$. Using again the eigenvalue-eigenvector decomposition
\[
\bo M_\tau = \bo U_\tau \bo D_\tau \bo U_\tau^\top,
\]
and separating the eigenvectors of $\bo M_\tau$ corresponding to non-zero and zero eigenvalues yields the $k_{\tau}\times p$ matrix $\bo U_{\tau,n}$ and the $(p-k_{\tau}) \times p$ matrix $\bo U_{\tau,s}$, where $k_\tau$ is the number of non-zero eigenvalues with  $k_\tau \leq k$. The unmixing matrix estimates are then $\bo W_{\tau,n} = \bo U_{\tau,n} \bo S_{0,T}(\bo x_t)^{-1/2}$ and $\bo W_{\tau,s} = \bo U_{\tau,s} \bo S_{0,T}(\bo x_t)^{-1/2}$. For this method to work, the different intervals are required to have different autocovariances in the intervals. As this method is aimed in detecting changes in the correlation structure we denote it as SSAcor.

\subsection{Comparison of SSAsir, SSAsave and SSAcor}
\label{sec:CompEx}

\begin{figure}[ht]
  \center
  \includegraphics[width=0.8\textwidth]{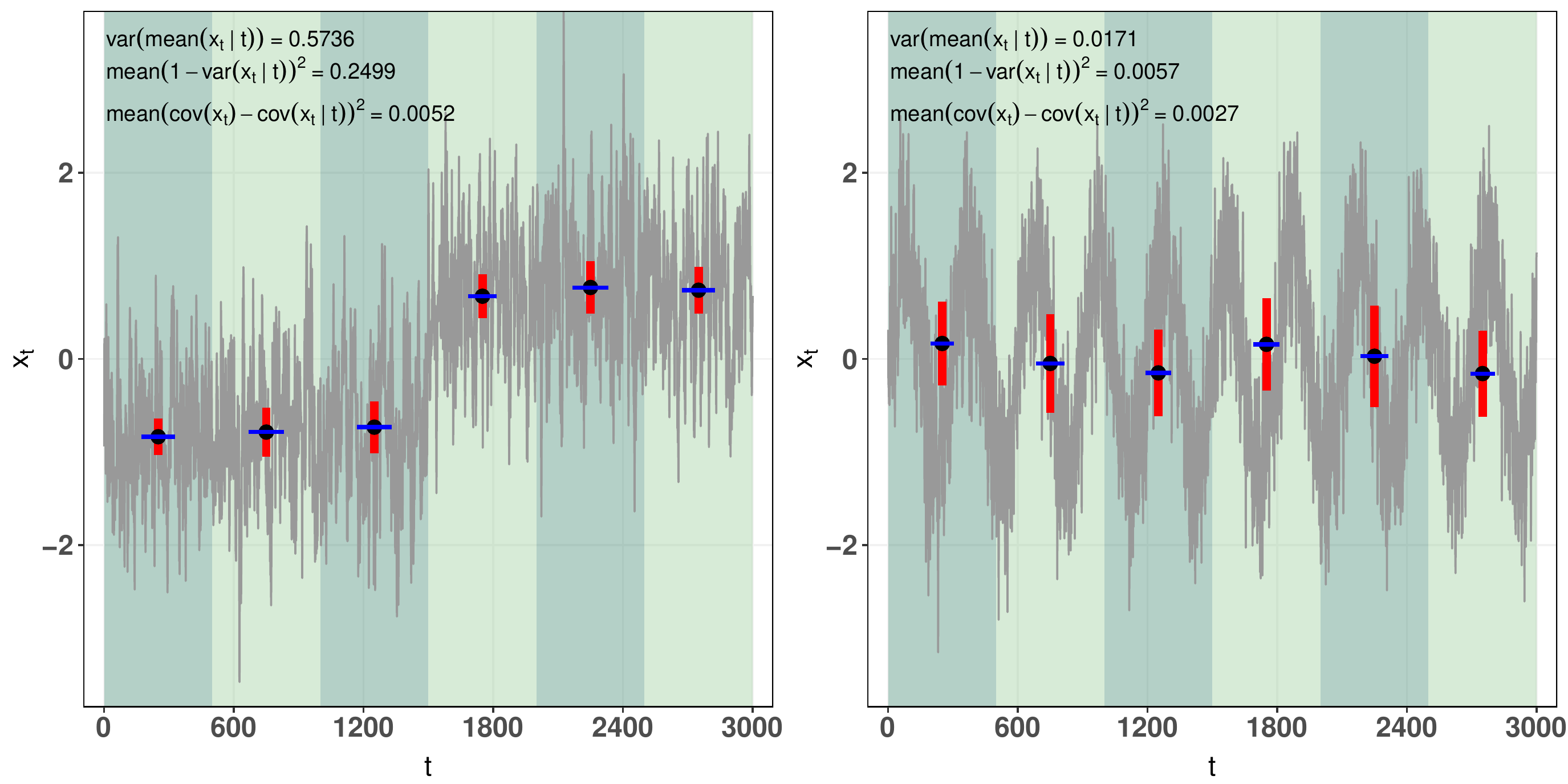}\\
  \caption{Visualization of SSAsir, SSAsave and SSAcor for a component with nonstationarity in mean (left panel trend, right panel seasonality). The black dots represent the interval mean, the height of vertical red bar the interval variance and the width of blue horizontal bar the interval autocovariance.}\label{Vis1}
\end{figure}

\begin{figure}[ht]
  \center
  \includegraphics[width=0.9\textwidth]{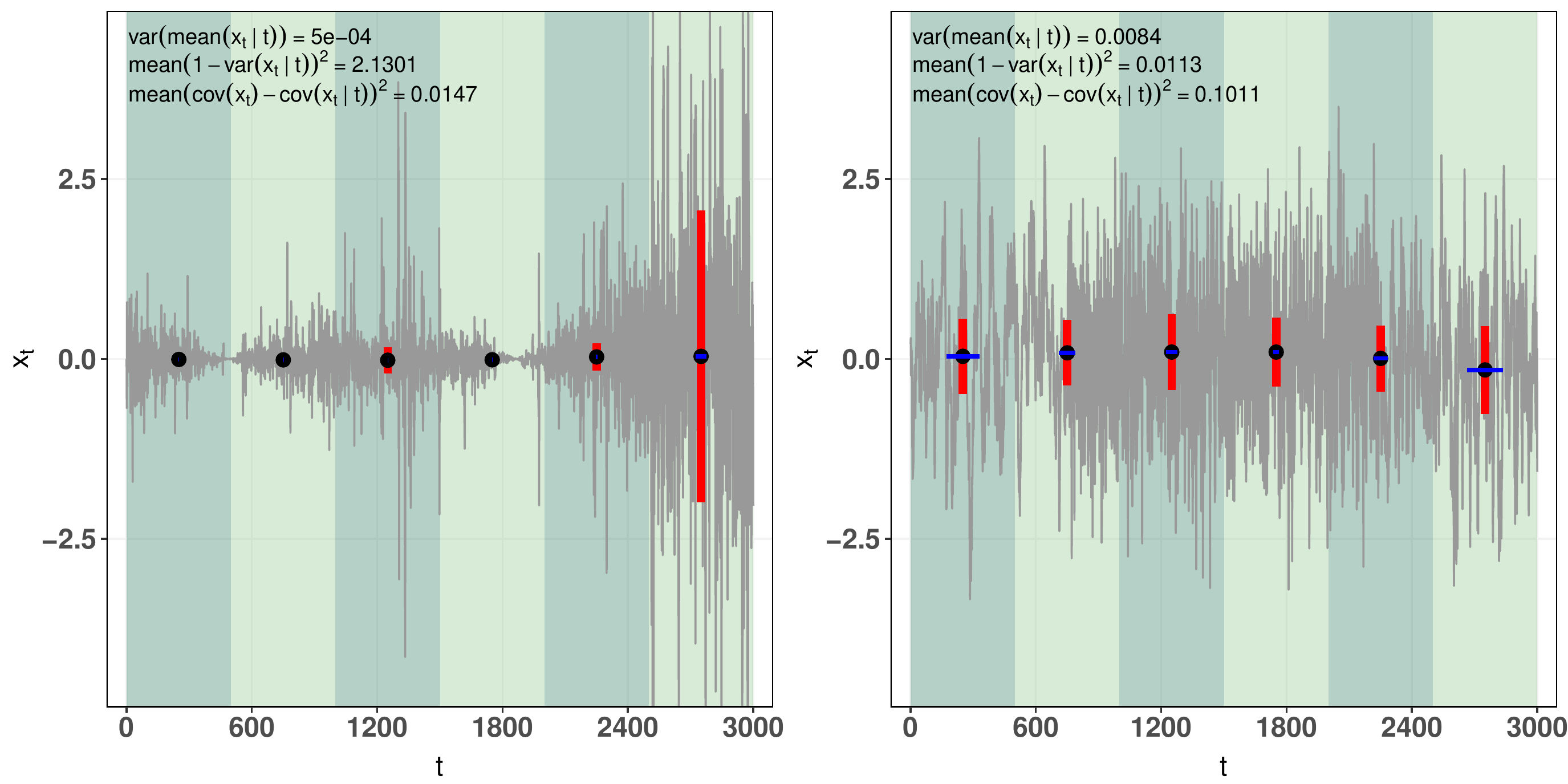}\\
  \caption{Visualization of SSAsir, SSAsave and SSAcor for a component with nonstationarity variance (left panel) and for a component with nonstationary correlation structure (right panel). The black dots represent the interval mean, the height of vertical red bar the interval variance and the width of blue horizontal bar the interval autocovariance.}\label{Vis2}
\end{figure}

In this section we visualize how SSAsir, SSAsave and SSAcor work and in which situations they fail. For that purpose we plot four nonstationary time series of length 3000 which are all standardized to have mean zero and unit variance. We consider $K=6$ equal-sized intervals and compute for each slice the mean, the variance and the autocovariance, and visualize these quantities. The three quantities are illustrated in Figures~\ref{Vis1} and~\ref{Vis2} using dots, vertical bars and horizontal bars, respectively. The methods now detect nonstationarities when there exists variation between the means (SSAsir), when the variances differ from 1 (SSA) which means that the intervals must have different variances, and when the autocovariance differs from the global autocovariance which means that they also need to be different. By combining the values computed at each slice, we then obtain the global statistics of interest for each time series and for each method. These statistics are reported in the top-left corners of the figures and for the methods to work the values need to be non-zero.  

Figure~\ref{Vis1} shows two different types of nonstationarities in mean. The left panel has a mean level shift and the right panel has a strong seasonal pattern. For the mean level shift the mean values are clearly non-constant but also the variances in the intervals differ. Thus in this case both SSAsir and SSAsave work though the statistic of SSAsir is much larger than that of SSAsave. Notice that SSAcor does not work at all in this case. In case of the seasonal time series the example shows that if the intervals are chosen badly, as is the case here, none of the methods work despite the clear nonstationarity. In~\ref{App:1}, Figure~\ref{VisA1} shows the same examples as in this section with $K=6$ intervals with unequal sizes. This demonstrates that SSAsir and SSAsave can perform well in the case of seasonal time series and the performance depends on the chosen intervals. 

In the left panel of Figure~\ref{Vis2} the time series has a constant mean but changes in variance. As expected SSAsir cannot detect this type of nonstationarity as only varying variances in the intervals are clearly visible. Again SSAcor does not work here. The right panel of Figure~\ref{Vis2} on the other hand has a constant mean and a constant variance but the correlation structure changes twice. This is also visible in the figure where for the first time the horizontal correlation bars appear to have clearly different lengths. However, the overall measure of nonstationarity used by SSAcor is still small indicating that detection of nonstationarity in correlation is quite difficult. As shown in Figure~\ref{VisA2} in the Appendix having intervals of unequal sizes does not give higher measure on nonstationarity for SSAcor. To conclude, there always seems to be cases where one of the methods is better than the other making approaches that combine the three methods a natural approach. 

\subsection{Combination of methods}
\label{sec:comb}

The methods described above can be combined in order to detect all three types of nonstationarities. Notice first that analytic SSA (ASSA) as proposed in \cite{Haraetal:2010} tries to recover the unmixing matrix $\bo W$ so that the covariances of the first and the second moments across intervals is minimized. To be more exact, the method combines the search for components that are nonstationary in mean and variance by using the eigenvalue-eigenvector decomposition of   
\[
\bo M_{ASSA} = \frac{1}{T}\sum_{i=1}^K \left\{\bo m_{T_i}(\bo y_t) \bo m_{T_i}(\bo y_t)^\top + \frac{1}{2} \bo S_{0,T_i}(\bo y_t) \bo S_{0,T_i}(\bo y_t)^T  \right\} -\frac{1}{2} \bo I_p,
\]
and proceeding as in the previous sections. The main drawback here is that the changes in autocorrelation are not necessarily detected. Furthermore there are no tools to identify what kind of nonstationarities the components exhibit.

Therefore, we suggest another approach which combines the three nonstationarity measures~(\ref{eq:mean}),~(\ref{eq:var}) and~(\ref{eq:cor}).  Denote from now on $\bo M_m= \bo M_1$, $\bo M_v= \bo M_2$ and $\bo M_\tau= \bo M_3$. Our suggestion is to turn the problem into a joint diagonalization problem, which means that we search for an orthogonal $p\times p$ matrix $\bo U_c$ which minimizes  
\[
\sum_i ||\off(\bo U_c^\top \bo M_i \bo U_c )||^2,
\]
or, equivalently, maximizes
\begin{equation}
\label{eq:jointdiag}
\sum_i ||\diag(\bo U_c^\top \bo M_i \bo U_c )||^2.
\end{equation}
Here $||\bo A||$ is the matrix (Frobenius) norm, $\diag(\bo A)$ is a $p\times p$ diagonal matrix with the diagonal elements as in $\bo A$ and $\off(\bo A) = \bo A - \diag(\bo A)$. Notice that naturally only a subset of the matrices or some additional matrices $\bo M_\tau$ with various lags $\tau$ can be added in the objective function, but the principle of the estimation procedure remains the same. Several algorithms for approximate joint diagonalization in~(\ref{eq:jointdiag}) exist in the literature. The most popular one based on Givens rotations is proposed in \cite{Clarkson:1988}. 

Based on $\bo U_c$ one can now compute $\bo D_i = \diag(\bo U_c^\top \bo M_i \bo U_c )=\diag(d_{i,1},\ldots, d_{i,p})$ and collect all those rows of $\bo U_c$, where $\sum_i d_{ij} \neq 0$, to a $k_c \times p$ matrix $\bo U_{c,n}$. The rest of the rows are then collected to a $(p-k_c) \times p$ matrix $\bo U_{c,s}$. Here the individual value $d_{i,j}$ indicates whether the $j$th component is nonstationary with respect to $\bo M_i$. Such classification of the components is not possible when, for example, methods such as ASSA are applied. The final transformation matrices for nonstationary and stationary components are then $\bo W_{c,n} = \bo U_{c,n} \bo S_{0,T}(\bo x_t)^{-1/2}$ and $\bo W_{c,s} = \bo U_{c,s} \bo S_{0,T}(\bo x_t)^{-1/2}$. In the following this method is referred as SSAcomb.


\section{Practical issues and identifiability of components}
\label{sec:practical}

There are several practical considerations still worth pointing out. Clearly the finite sample eigenvalues of matrices $\bo M$ corresponding to the stationary components will be never exactly zero. Thus the value of $k$ must be chosen based on a cut-off value or graphically. Furthermore, it is obvious that the choice of the number of intervals $K$ and how they are divided is crucial in this framework. The impact of the number of intervals will be illustrated in the simulation studies in Section~\ref{sec:simulations}. Ideally, given the number of intervals, the cut points of the intervals should be such that the quantities of interest are as different as possible. However, as finding optimal cut points is in practice difficult, the intervals should, in our opinion, be at least of different length so that possible seasonality effects are easier to detect and issues similar to those seen in Section~\ref{sec:CompEx} do not occur.

For all methods discussed above, the unmixing matrix $\bo W$ has the two parts $\bo W_{n} = \bo U_{n} \bo S_{0,T}(\bo x_t)^{-1/2}$ and $\bo W_{s} = \bo U_{s} \bo S_{0,T}(\bo x_t)^{-1/2}$ specifying nonstationary and stationary subspaces, respectively, where matrices $\bo U_n$ and $\bo U_s$ depend on the method used. Let now $\bo x_t^* = \bo B \bo x_t $ denote an affine transformation of $\bo x_t$ with full-rank $p\times p$ matrix $\bo B$. We then denote the unmixing matrix based on $\bo x_t^*$ as $\bo W^*$. In the BSS literature (see for example \cite{MiettinenTaskinenNordhausenOja2015}) a BSS unmixing matrix is defined to be affine equivariant if $\bo W \bo x_t = \bo J \bo W^* \bo x_t^*$ for some diagonal matrix $\bo J$ which has $\pm 1$ on its diagonal. The affine equivariance thus means that the recovered components do not depend on the mixing matrix, except for their signs.

In the proposed SSA approach, the eigenvalues of $\bo M$ matrices provide a way of ordering nonstationary components (or pseudo-eigenvalues in the case of SSAcomb). Thus if the eigenvalues corresponding to the nonstationarity components are unique then $\bo U_n$ is well defined and we have $\bo W_n \bo x_t = \bo J \bo W_n^* \bo x_t^*$. However, for the stationary part all (pseudo-)eigenvalues are equal and thus $\bo U_s$ is not well defined and we actually have $\bo W_s \bo x_t = \bo O \bo W_n^* \bo x_t^*$ for some orthogonal matrix $\bo O$. However, as for finite data the eigenvalues are usually all distinct, the finite sample version will be affine equivariant. Note however that $\bo W_s \bo x_t$ is not necessarily equivalent to $\bo s_t$ and $\bo W_n \bo x_t$ is not necessarily equivalent to $\bo n_t$ as both $\bo s_t$ and $\bo n_t$ have the  ambiguities mentioned above and only the two subspaces are well defined. This is quite similar to a non-Gaussian subspace analysis (NGCA) \cite{BlanchardKawanabeSugiyamaSpokoinyMuller2006} - a BSS method for iid data that divides data into a Gaussian part and a non-Gaussian part. See also the supplementary material of~\cite{BunauMeineckeKiralyMuller:2009} for a detailed discussion.  However, it is often desirable to have models where the interesting components are well defined (up to some minor identifiability issues such as their signs). In this work we do not consider a full analysis of the identifiability issues, just mention some possibilities. The stationary components are identifiable when the components follow a second-order source separation (SOS) model or a stationary independent time series model. For details see for example \cite{PanMatilainenTaskinenNordhausen2021}. In that case methods such as AMUSE \cite{TongSoonHuangLiu1990,MiettinenNordhausenOjaTaskinen2012}, SOBI \cite{BelouchraniAbedMeraimCardosoMoulines1997,Miettinenetal:2016}, gSOBI \cite{MiettinenMatilainenNordhausenTaskinen2017} or gJADE \cite{MatilainenNordhausenOja2015} can be applied to the identified stationary subspace. For the nonstationary subspace at least the following assumptions allow the estimation of the components. If all nonstationary components are independent and the marginal distributions are non-Gaussian, independent component analysis (ICA) methods can be used. For an overview of several ICA methods, see \cite{ComonJutten:2010,NordhausenOja2018}. Another possibility is to assume a nonstationary source separation (NSS) model for the nonstationary components, that is, to assume that the mean is stationary, but each component has nonstationary variances which change in different patterns. This is for example a reasonable assumption for audio data. Interestingly, most NSS methods also divide the time series into intervals and jointly diagonalize statistics based upon them. See for example \cite{ChoiCichocki:2000,Nordhausen2014} for more details on NSS methods. In this context SSAsave seems to be a natural SSA method. 

So far we have assumed that the dimensions of the two subspaces are known, as it is assumed in almost all of the SSA literature mentioned above. This is naturally a very unnatural assumption. As long as inferential tools are missing to estimate $k$, the screeplot of the (pseudo-)eigenvalues might give some indication of the choice.
\begin{figure}[ht]
  \center
  \includegraphics[width=0.7\textwidth]{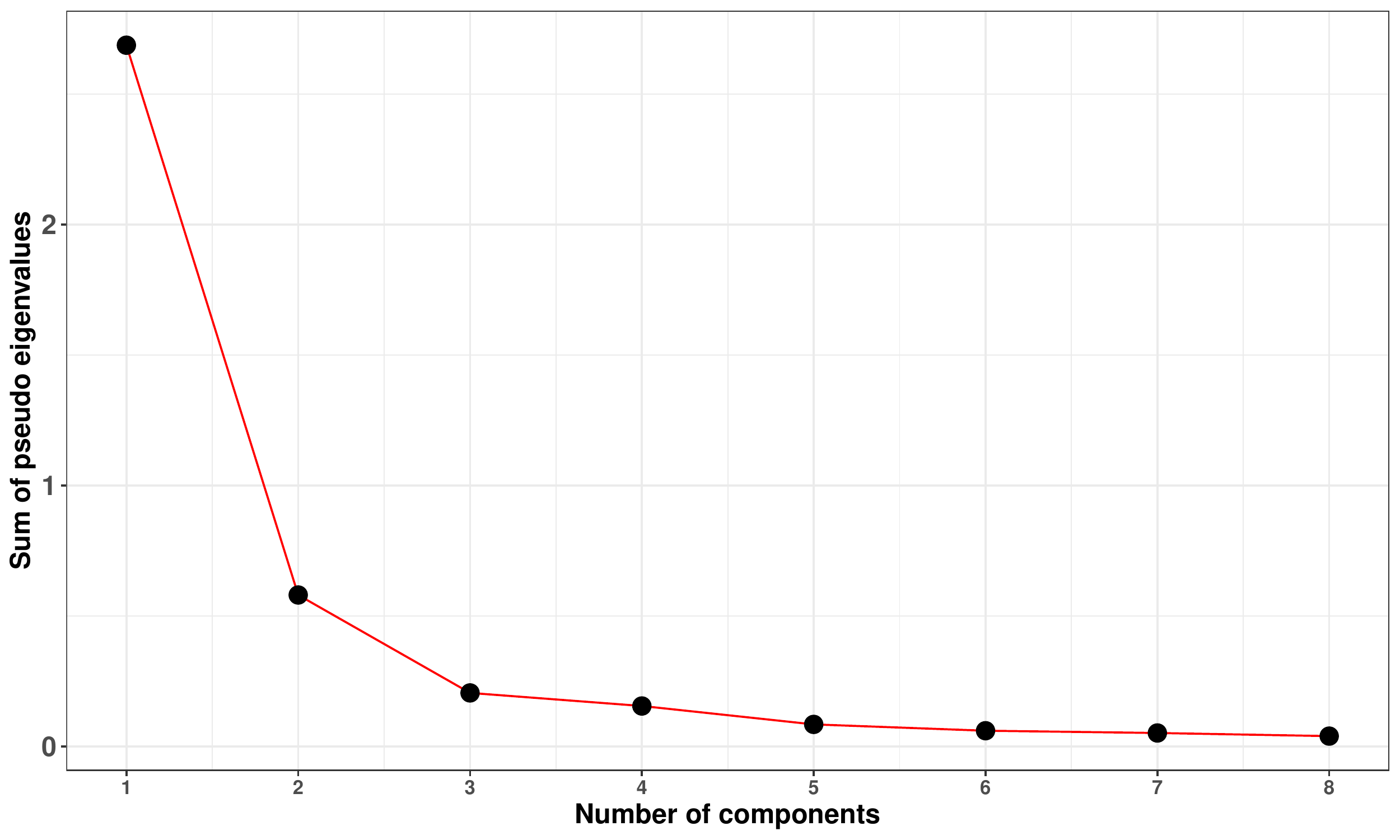}\\
  \caption{Screeplot based on the column sums of the pseudo-eigenvalues from Table~\ref{tab:ddd}.}\label{screeplot1}
\end{figure}
In the following we apply SSAcomb based on six equal sized intervals for an eight-variate time series with $T=4000$. The time series are simulated according to Setting 4 in the following section.
\begin{table}[ht]
\center
\begin{tabular}{rrrrrrrrr}
  \hline
 & $d_1$ & $d_2$ & $d_3$ & $d_4$ & $d_5$ & $d_6$ & $d_7$ & $d_8$ \\ 
  \hline
$\bo M_m$ & 0.00 & 0.53 & 0.00 & 0.02 & 0.01 & 0.00 & 0.00 & 0.00 \\ 
$\bo M_v$ & 2.67 & 0.03 & 0.03 & 0.07 & 0.03 & 0.03 & 0.02 & 0.02 \\ 
$\bo M_\tau$ & 0.02 & 0.02 & 0.18 & 0.07 & 0.04 & 0.03 & 0.02 & 0.02 \\ 
   \hline
\end{tabular}
\caption{Pseudo-eigenvalues of SSAcomb for a simulated time series.}
\label{tab:ddd}
\end{table}
Table~\ref{tab:ddd} contains the pseudo-eigenvalues and the screeplot shown in Figure~\ref{screeplot1} is based on their column sums. Based on the screeplot one could think of 3 or 4 nonstationary components. Taking then into account the underlying pseudo-eigenvalues it is clear that the first component is nonstationary according to $\bo M_v$ only indicating nonstationarity in variance. The second component is nonstationarity according to $\bo M_m$ and the third component according to $\bo M_\tau$. The fourth component actually has small values for all three $\bo M$-matrices and therefore might actually not be really nonstationary. The true value of $k$ is indeed three in this case. 

\section{Simulations and an example}
\label{sec:simulations}

In this section we compare the methods discussed above using simulation studies. The method that combines the search for all three types of nonstationarities is also illustrated by an example.

\subsection{Simulations}

As pointed out above, it depends on the purpose of the analysis whether the stationary subspace or the nonstationary subspace is of interest. The goal in this simulation study is therefore to evaluate how well the different methods can estimate the two subspaces under the assumption that the dimension of the subspace is known, as it is assumed in most of the SSA references mentioned above. Thus, based on the sample eigenvalues of the methods, the unmixing matrix estimates can be decomposed into $\hat \bo W_s$ and $\hat \bo W_n$ with dimensions $(p-k) \times p$ and $k \times p$, respectively.

For comparing the methods, we need a performance measure that takes into account the fact that the subspaces are only specified up to rotations. We therefore compute the projection matrices
$$
\hat \bo P_s = \hat \bo W_s (\hat \bo W_s^\top \hat \bo W_s)^{-1} \hat \bo W_s^\top \quad \mbox{and} \quad 
\hat \bo P_n = \hat \bo W_n (\hat \bo W_n^\top \hat \bo W_n)^{-1} \hat \bo W_n^\top
$$
and will compare them with the corresponding true projection matrices $\bo P_s$ and $\bo P_n$ by computing the squared distances
$$
D_s^2=\frac{1}{{2}} \| \bo P_s- \hat \bo P_s\|^2
\quad \mbox{and} \quad
D_n^2=\frac{1}{{2}} \| \bo P_n- \hat \bo P_n\|^2,
$$
which can take values in $[0,\min\{k,p-k\}]$, with zero indicating a perfect recovery of the subspace. For more details about this performance criterion, see \cite{CroneCrosby1995,LiskiNordhausenOjaRuizGazen2016}.

All following simulations were done with R \cite{R} using the packages JADE \cite{JADE_package} and LDRtools \cite{LDRtools}. In all simulation settings the observed time points are $1, \dots, T$, where $T$ varies from 1000 to 32000, and $p=8$ with five stationary components $s_{1,t},\ldots,s_{5,t}$ and three nonstationary components $n_{1,t}$,$n_{2,t}$ and $n_{3,t}$.   Most components are based on moving average (MA) processes, autoregressive (AR) processes and autoregressive moving average (ARMA) processes. However, inspired by \cite{PatileaRaissi2014} we consider also time varying variance (TV-VAR) processes $x_t$ with parameters $\alpha$ and $\beta$ of the form
$x_t = \tilde h_t \epsilon_t$ where $\tilde h_t^2 = h_t^2 + \alpha x_{t-1}^2$ where $\epsilon$ are iid $N(0,1)$, $x_0=0$ and  $h_t =  10 - 10 \sin(\beta \pi \ t/T + \pi/6)  (1 + t/T)$.
Similarly we consider also time varying autoregressive processes of order 1 (TV-AR1) where $x_t= a_t x_{t-1} + \epsilon_t$ with $x_0=0$, $\epsilon_t$ is iid $N(0, \sigma_2)$ and $a_t = 0.5 \cos(2\pi \ t/T)$. 

\begin{figure}[ht]
  \center
  \includegraphics[width=0.8\textwidth]{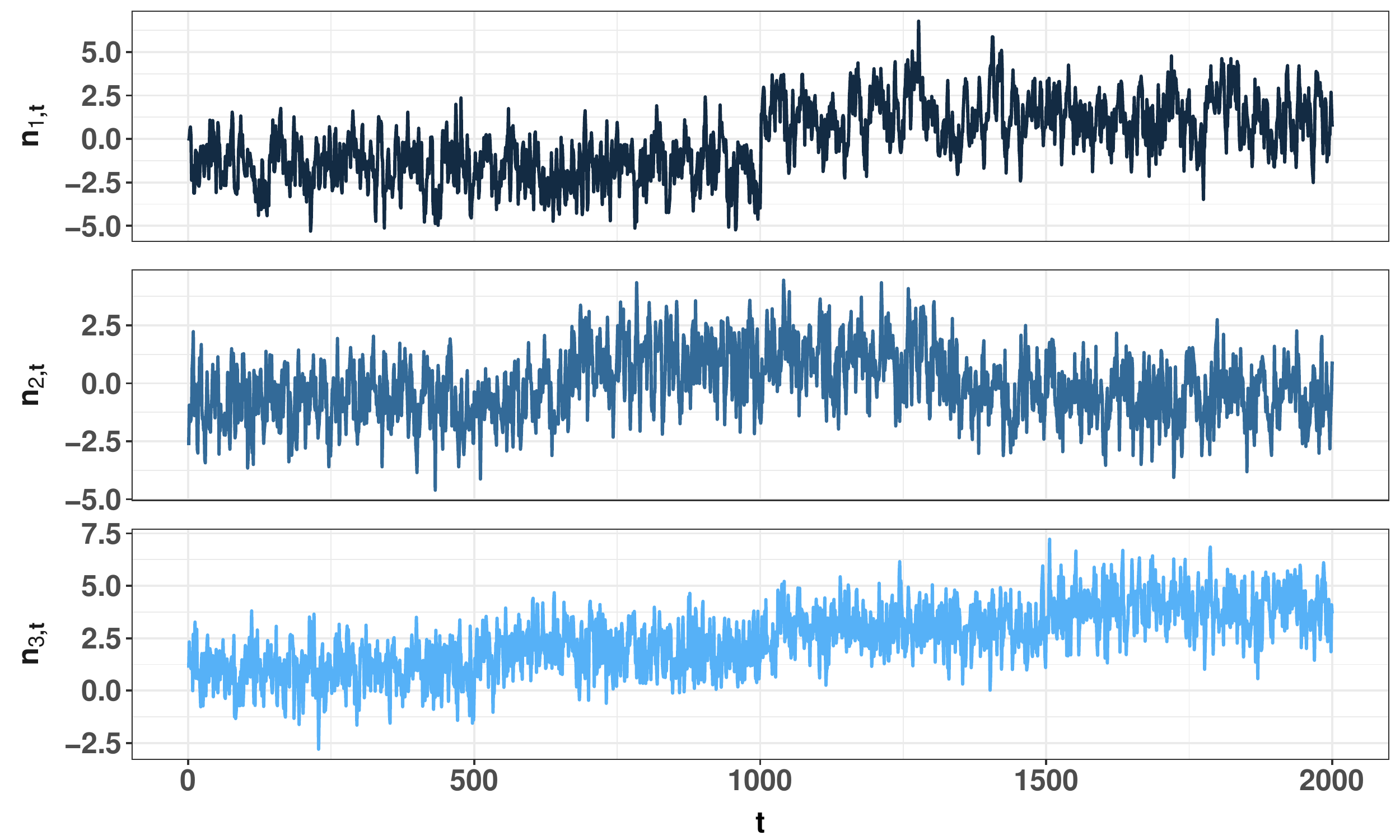}\\
  \caption{Example of the nonstationary components in Setting~1 for $T=2000$.}\label{ExNSset1}
\end{figure}

\begin{figure}
  \center
  \includegraphics[width=0.9\textwidth]{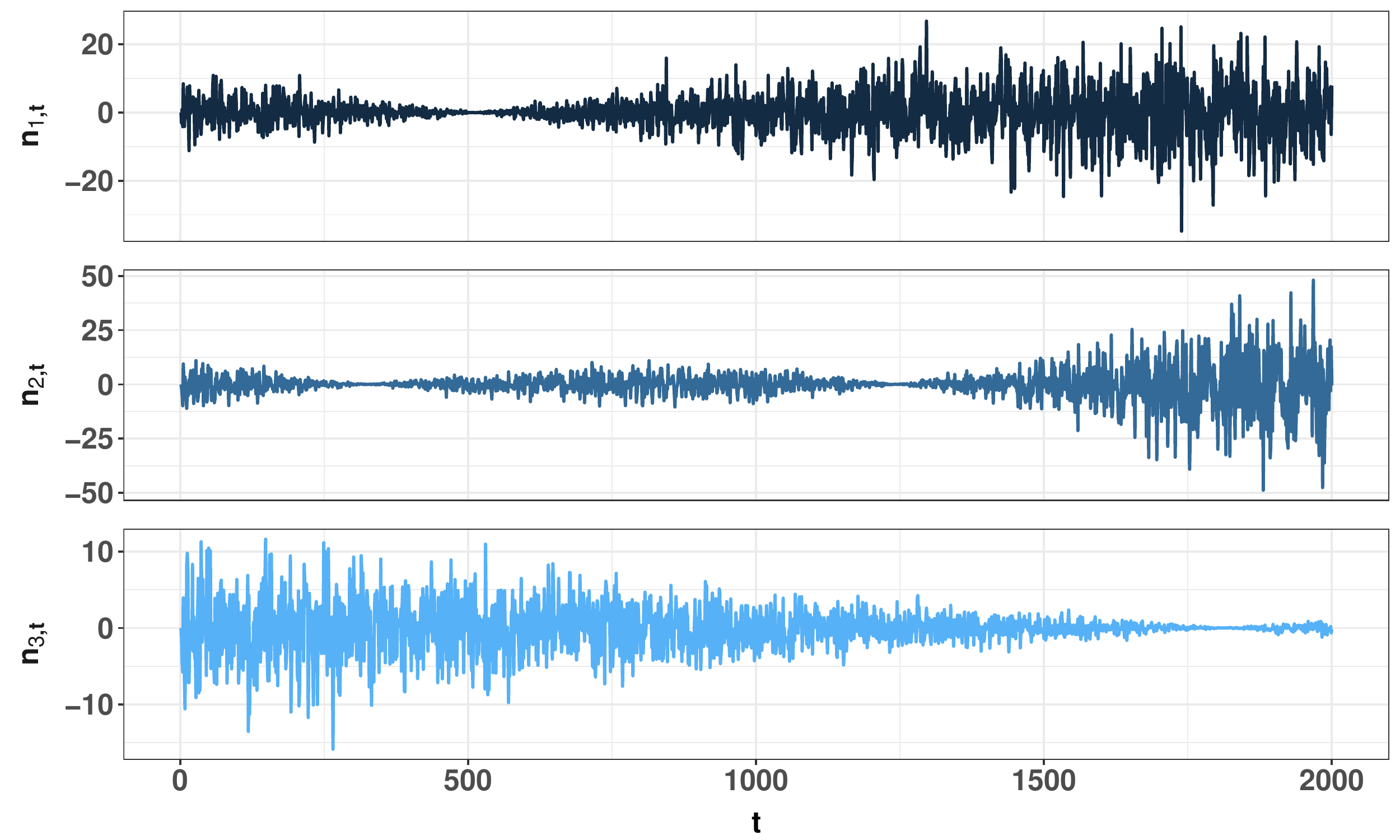}\\
  \caption{Example of the nonstationary components in Setting~2 for $T=2000$.}\label{ExNSset2}
\end{figure}

\begin{figure}
  \center
  \includegraphics[width=0.9\textwidth]{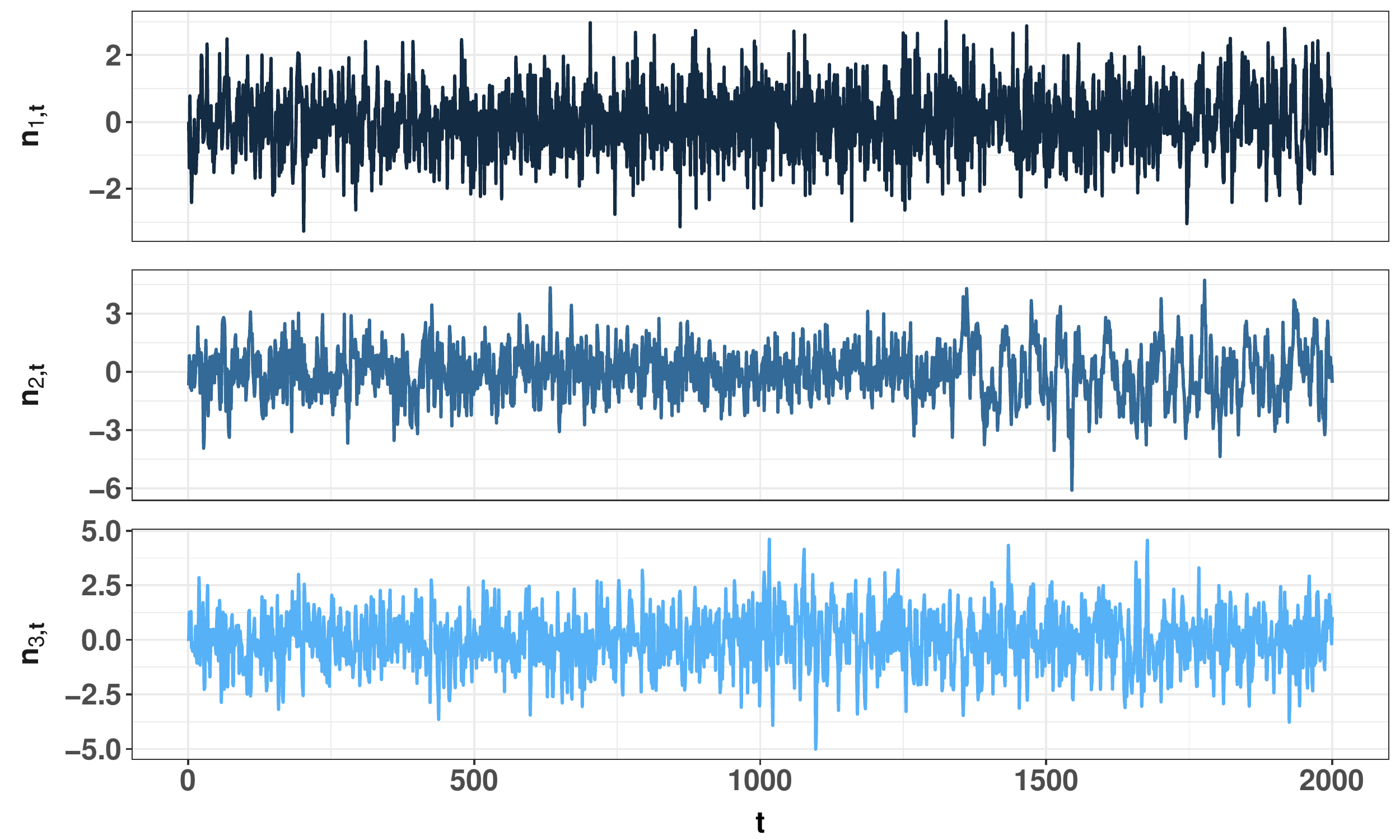}\\
  \caption{Example of the nonstationary components in Setting~3 for $T=2000$.}\label{ExNSset3}
\end{figure}

The four 8-variate settings under consideration are then:
\begin{description}
\item[Setting 1:] $s_{1,t}$ is a MA(0.72, 0.24) process, $s_{2,t}$ is an AR(0.34, 0.27, 0.18) process, $s_{3,t}$ is an ARMA(0.34, 0.27, 0.18; 0.72, 0.15) process, $s_{4,t}$ is an AR(0.11, 0.58) process and $s_{5,t}$ is an MA(0.78) process. $n_{1,t} = y_t + \mu_t$ where $y_t$ is an AR(0.7) process and $\mu_t = -1.52$ if $t \leq \left \lfloor{T/2}\right \rfloor$ and otherwise $1.38$. $n_{2,t} = y_t + \mu_t$ where $y_t$ is an AR(0.5) process and $\mu_t = -0.75$ if $t \leq \left \lfloor{T/3}\right \rfloor$, 0.84 if $t \in [\left \lfloor{T/3}\right \rfloor + 1, 2 \left \lfloor{T/3}\right \rfloor]$ 
and otherwise $-0.45$. 
\item[Setting 2:] $s_{1,t}$ is a MA(0.72) process, $s_{2,t}$ is a MA(0.34) process, $s_{3,t}$ is a MA(0.72, 0.15), $s_{4,t}$ is an MA(0.11, 0.58) process and $s_{5,t}$ is a MA(0.34, 0.27, 0.18) process. $n_{1,t}$ is TV-VAR(0.2,0.5), $n_{2,t}$ is TV-VAR(0.1,1) and $n_{3,t}$ is TV-VAR(0.05,0.01). 
\item[Setting 3:] $s_{1,t}$ is an ARMA(0.14, 0.45; 0.72, 0.24) process, $s_{2,t}$ is an AR(0.34, 0.27, 0.18) process, $s_{3,t}$ is an ARMA(0.34, 0.27, 0.18; 0.72, 0.15) process, $s_{4,t}$ is an AR(0.11,0.58) process and $s_{5,t}$ is an AR(0.1, 0.1, 0.1, 0.1, 0.1) process.
$n_{1,t}$ is a TV-AR process with $\sigma_2 =  0.8649$, $n_{2,t}$ consists of three independent blocks, i.e., for $t=1$ to  $t=\left \lfloor{T/3}\right \rfloor$ an AR(0.5) process with $\sigma^2=1$, for $t=\left\lfloor{T/3}\right \rfloor+1$ to $t=2\left\lfloor{T/3}\right \rfloor$ an AR(0.2) process with $\sigma^2=1.6384$ and for $t=2\left\lfloor{T/3}\right \rfloor+1$ to $t=T$ an AR(0.8) process with $\sigma^2=0.2304$. $n_{3,t}$ consists of two independent blocks, i.e., for times $t=1$ to  $t=\left \lfloor{T/2}\right \rfloor$ a  MA(0.5) process with $\sigma^2=1$ and for $t=2\left \lfloor{T/2}\right \rfloor+1$ to $t=T$ a MA(0.9, 0.17) process with $\sigma^2=0.4624$
\item[Setting 4:] Has the same stationary components as used in Setting 3 and $n_{1,t}$ corresponds to $n_{1,t}$ in Setting 1,
$n_{2,t}$ corresponds to $n_{2,t}$ in Setting 2 and $n_{3,t}$ corresponds to $n_{1,t}$ in Setting 3.
\end{description}
 Thus, in Setting 1 the three nonstationary components have all nonstationary means, while in Setting 2 the means are all stationary but the variances are nonstationary. In Setting 3 the nonstationary information is in the correlation structure. Setting 4 uses then one nonstationary component from each of the previous settings meaning that here combining different approaches seems especially important. Examples of nonstationary components for Settings 1-3 with $T=2000$ are shown in Figures~\ref{ExNSset1}-\ref{ExNSset3} showing that the violations of nonstationarity can take quite different forms.

In the simulation study we simulated 2000 source sets with various sample sizes $T$ from all four settings and always used a random orthogonal mixing matrix as $\bo A$. Then for all methods we computed the distances between the true projections and the estimated projections when using $K=2,6,12$ equal-sized intervals. Notice that in the spirit of SIR, the number of slices for SSAsir should exceed $k=3$ and therefore SSAsir is not expected to work when $K=2$.

\begin{figure}[ht]
  \center
   \includegraphics[width=0.8\textwidth]{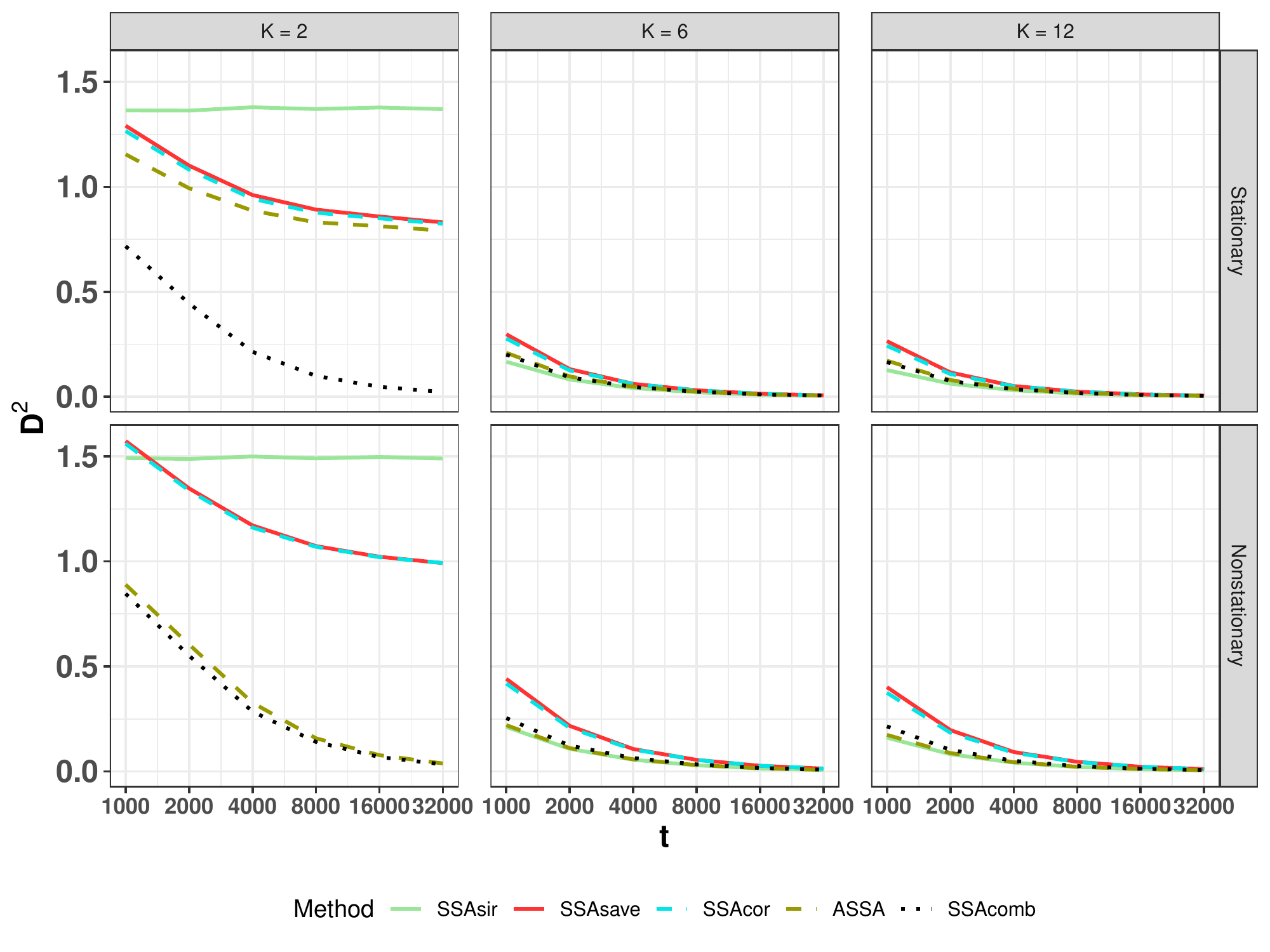}\\
  \caption{Average squared distances between the true and estimated subspaces in Setting~1.
  }\label{DistSset1}
\end{figure}

\begin{figure}[ht]
  \center
   \includegraphics[width=0.8\textwidth]{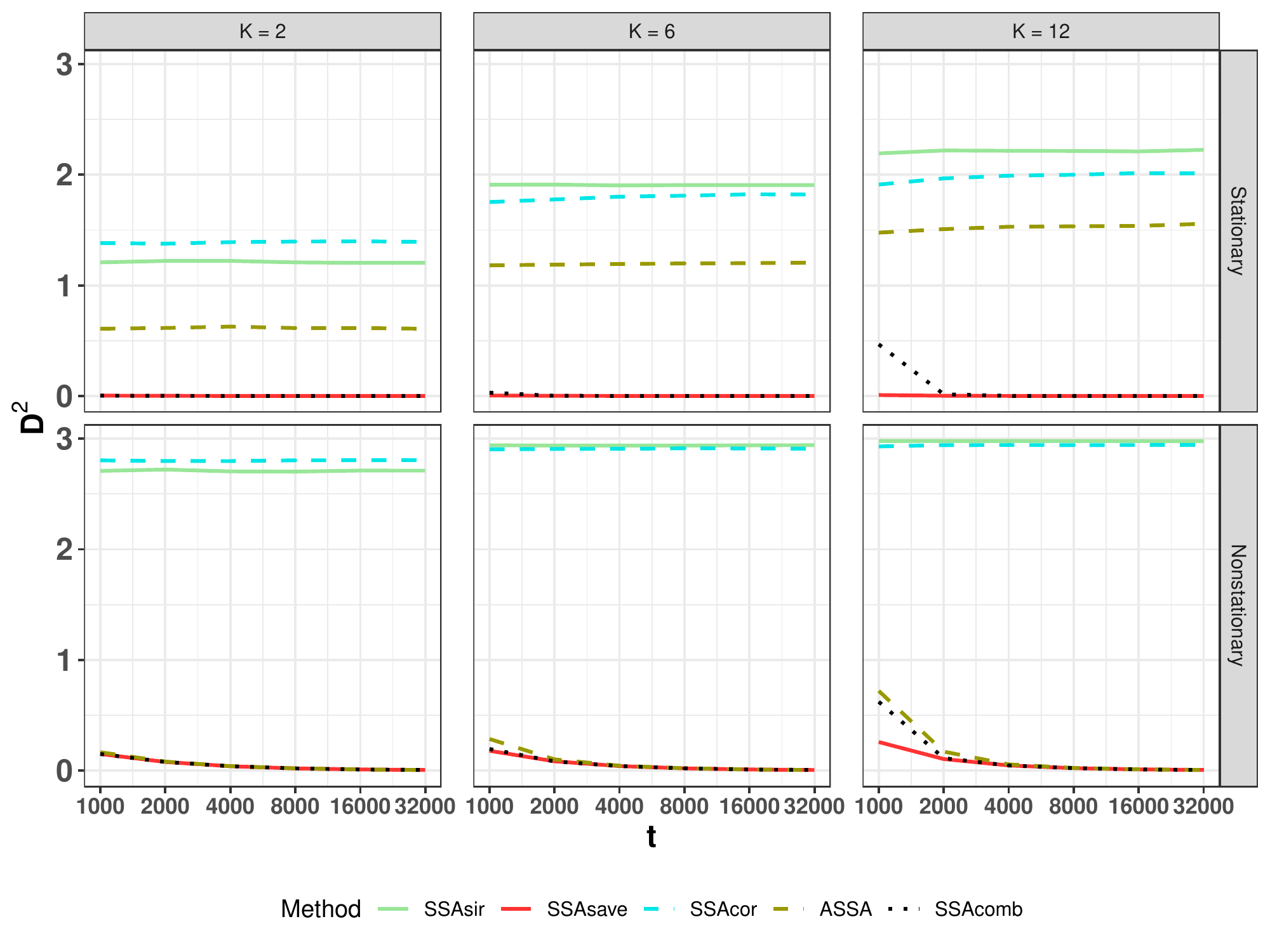}\\
  \caption{Average squared distances between the true and estimated subspaces in Setting~2.}\label{DistSset2}
\end{figure}

\begin{figure}[ht]
  \center
   \includegraphics[width=0.8\textwidth]{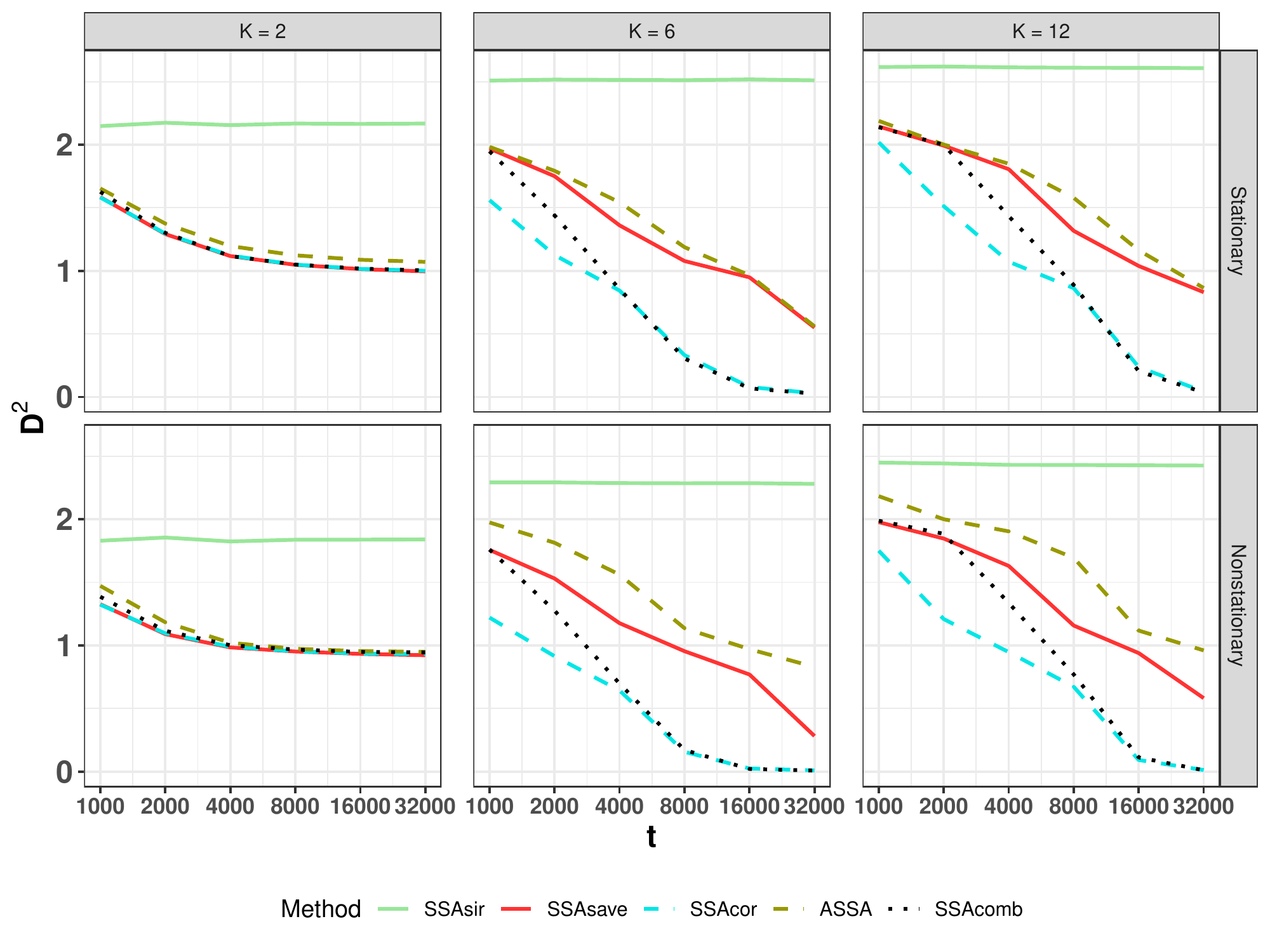}\\
  \caption{Average squared distances between the true and estimated subspaces in Setting~3.}\label{DistSset3}
\end{figure}

\begin{figure}[ht]
  \center
   \includegraphics[width=0.8\textwidth]{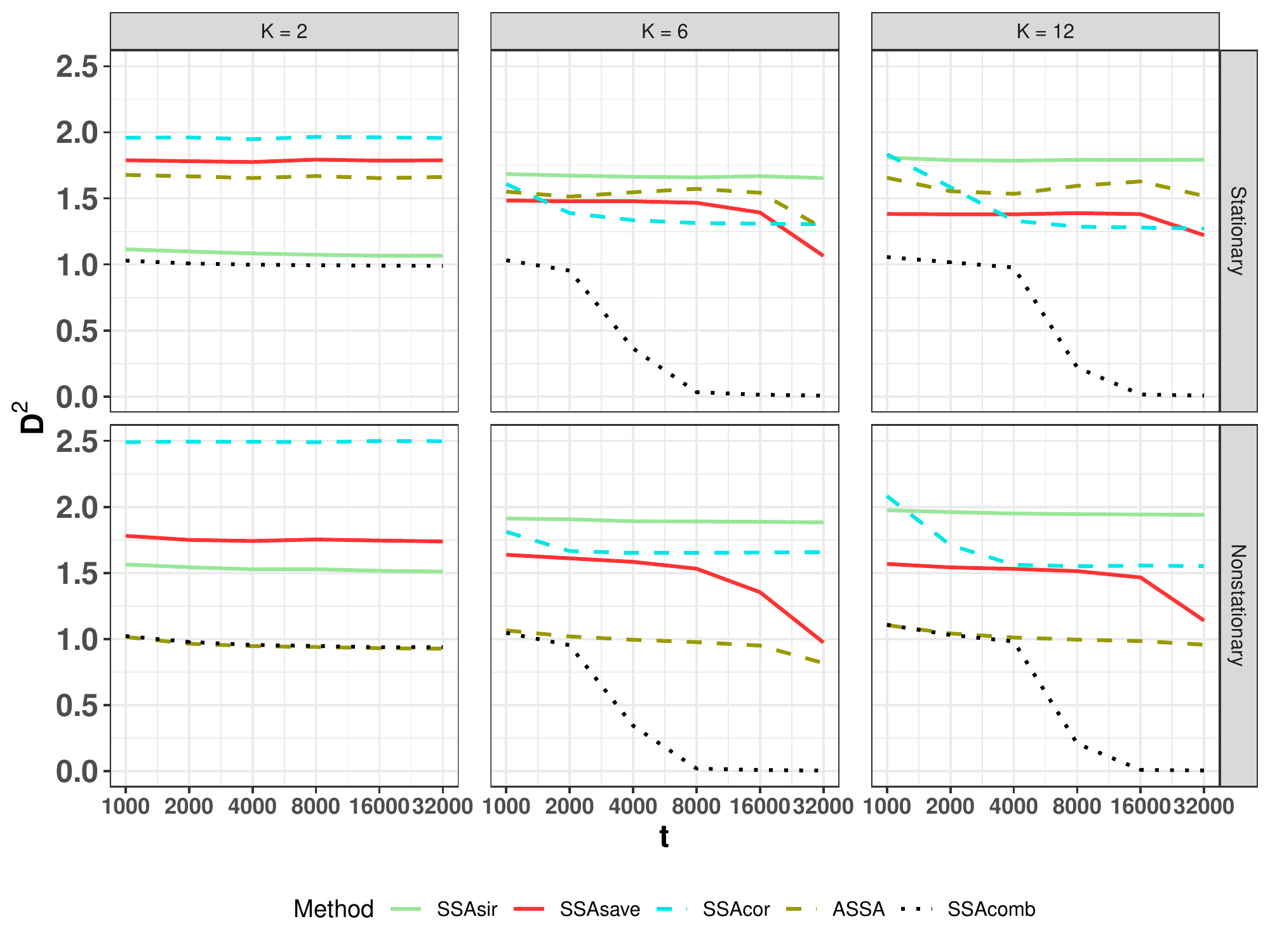}\\
  \caption{Average squared distances between the true and estimated subspaces in Setting~4.}\label{DistSset4}
\end{figure}

The average distances are shown in Figures~\ref{DistSset1}-\ref{DistSset4}. It is clearly visible that having only two intervals, that is $K=2$, is always the poorest choice. Although this was an expected result for SSAsir, it was not so obvious for the other methods. The differences between $K=6$ and $K=12$ decrease with increasing sample size. However for smaller sample sizes the choice $K=6$ is preferable. This confirms results familiar from NSS and SDR studies which show that it is important that the intervals contain enough observations to estimate the quantities of interest well enough.

As seen in Figure~\ref{DistSset1}, in case of Setting 1 all methods are able to estimate the subspaces when there are enough intervals. This is a bit surprising result for SSAcor. From the results based on Setting 2 in Figure~\ref{DistSset2} it is clear that SSAsir and SSAcor do not work at all. ASSA is unable to detect the stationary subspace but detects the nonstationary subspace quite well. SSAcomb works well despite containing matrices which, when used by themselves, fail completely. The results based on Setting 3 in Figure~\ref{DistSset3} indicate that SSAcor and SSAcomb perform clearly best while ASSA does worse than SSAsave and is therefore affected by the poor performance of SSAsir. Finally, as seen in Figure~\ref{DistSset4}, in Setting 4, where all different types of nonstationarities are present, it is clear that SSAcomb is the only method that can estimate the subspaces properly. It requires however a much larger sample size to detect all three nonstationary components. ASSA seems to be constantly missing one nonstationary component which is not surprising as it is not looking for nonstationarity in autocorrelation structures.

Thus, to conclude, based on the simulation studies SSAcomb seems to be a preferable method as it works quite well independently of the nature of the underlying stationarity violations. However, large sample sizes are preferable and the number of intervals $K$ plays a role in the performance. In the simulation studies above, a moderate number of $K=6$ seems as a reasonable choice.

\subsection{Example}

We applied the proposed SSAcomb method to a magnetoencephalographic (MEG) data of length $T=221710$ that was recorder using $p=102$ magnetometers at the Centre for Interdisciplinary Brain Research, Department of Psychology, University of Jyv\"askyl\"a. Magnetometers record brain activity indirectly by measuring changes in magnetic fields generated by electrical currents occurring in the brain. As the measurements are taken from top of the head, it is realistic to assume that the observed signals are mixtures of actual brain activity signals. The signals are also mixed with artifacts that occur due to external physiological factors such as moving the head, tensing muscles and blinking and moving eyes. Such mixing is also seen in Figure~\ref{MEGoriginal} in~\ref{App:2} where we plot ten first MEG-signals. The goal of the analysis is then to recover the brain activity of interest from the observed mixture.  Notice that KL-SSA was applied to EEG data analysis with the similar goal in mind in \cite{BunauMeineckeKiralyMuller:2009,Bunauetal:2010}. 

\begin{figure}[ht]
  \center
  \includegraphics[width=0.8\textwidth]{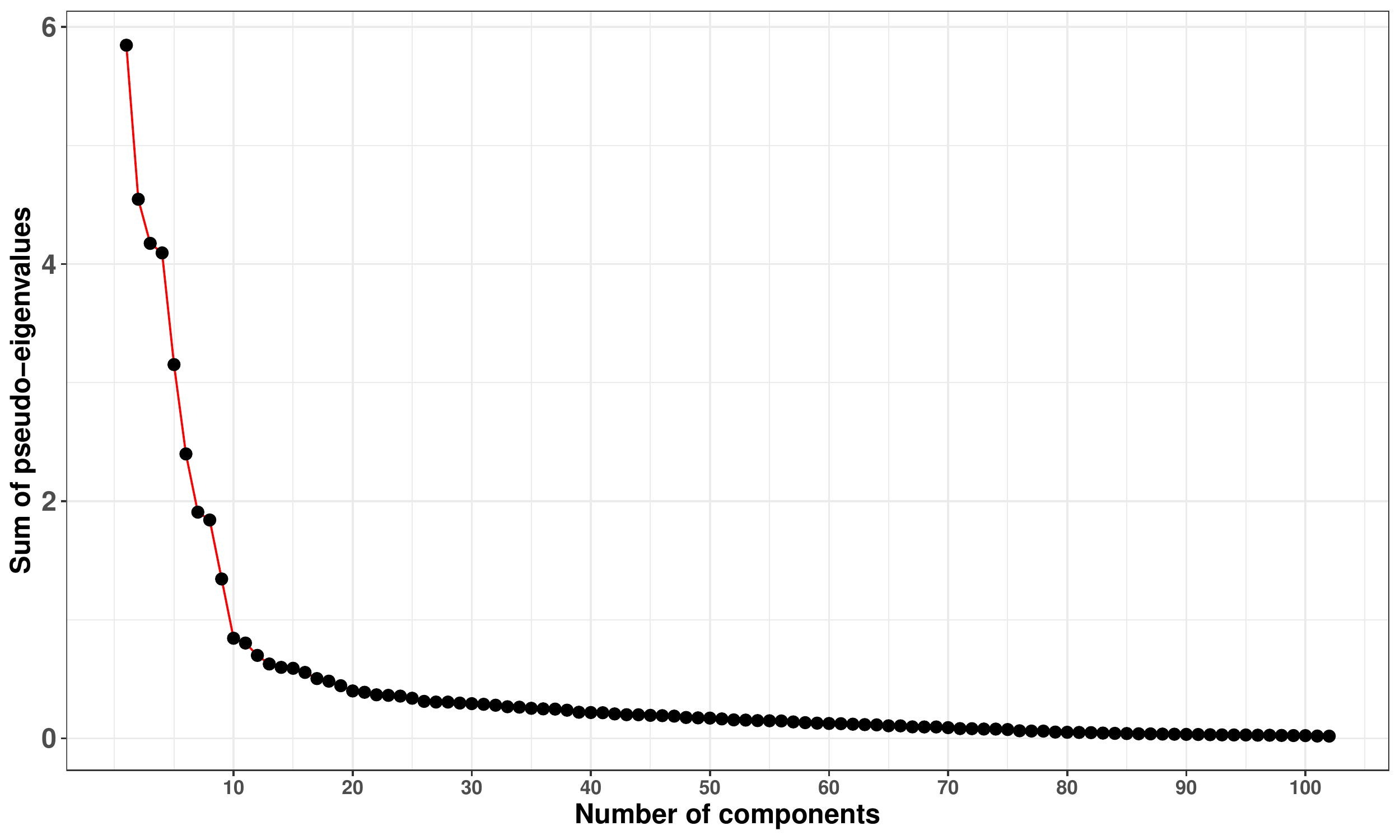}\\
  \caption{Screeplot of the sums of the pseudo-eigenvalues based on MEG data.}\label{screeplot2}
\end{figure}

Our interest was to see if SSAcomb can recover nonstationary components that possibly correspond to interesting brain activity or physiological signals. As the length of the data was pretty large, we applied SSAcomb with $K=12$, and plotted the screeplot of the sums of the pseudo-eigenvalues in Figure~\ref{screeplot2}. 

Based on the screeplot we note that the method is able to detect ten nonstationary components. We then list the ten pseudo-eigenvalues in Table~\ref{tab:MEGddd} and conclude that the components 1-5 and 8-10 were detected due to nonstationarity in variance and in autocorrelation. Component 6 has all three types of nonstationarities and component 7 is nonstationary with respect to the variance. 

\begin{table}[ht]
\centering
\begin{tabular}{rrrrrrrrrrr}
  \hline
 & $d_1$ & $d_2$ & $d_3$ & $d_4$ & $d_5$ & $d_6$ & $d_7$ & $d_8$ & $d_9$ & $d_{10}$\\ 
  \hline
$\bo M_m$ &   0.08 & 0.06 & 0.00 & 0.01 & 0.01 & 0.41 & 0.00 & 0.00 & 0.03 & 0.00 \\ 
$\bo M_v$ &   2.90 & 2.26 & 2.47 & 2.07 & 1.59 & 1.03 & 1.89 & 1.18 & 0.68 & 0.45 \\ 
$\bo M_\tau$ & 2.87 & 2.22 & 1.70 & 2.02 & 1.55 & 0.96 & 0.02 & 0.66 & 0.64 & 0.39 \\ 
   \hline
\end{tabular}
\caption{Pseudo-eigenvalues of SSAcomb based on the MEG data.}
\label{tab:MEGddd}
\end{table}

\begin{figure}[ht]
  \center
  \includegraphics[width=0.8\textwidth]{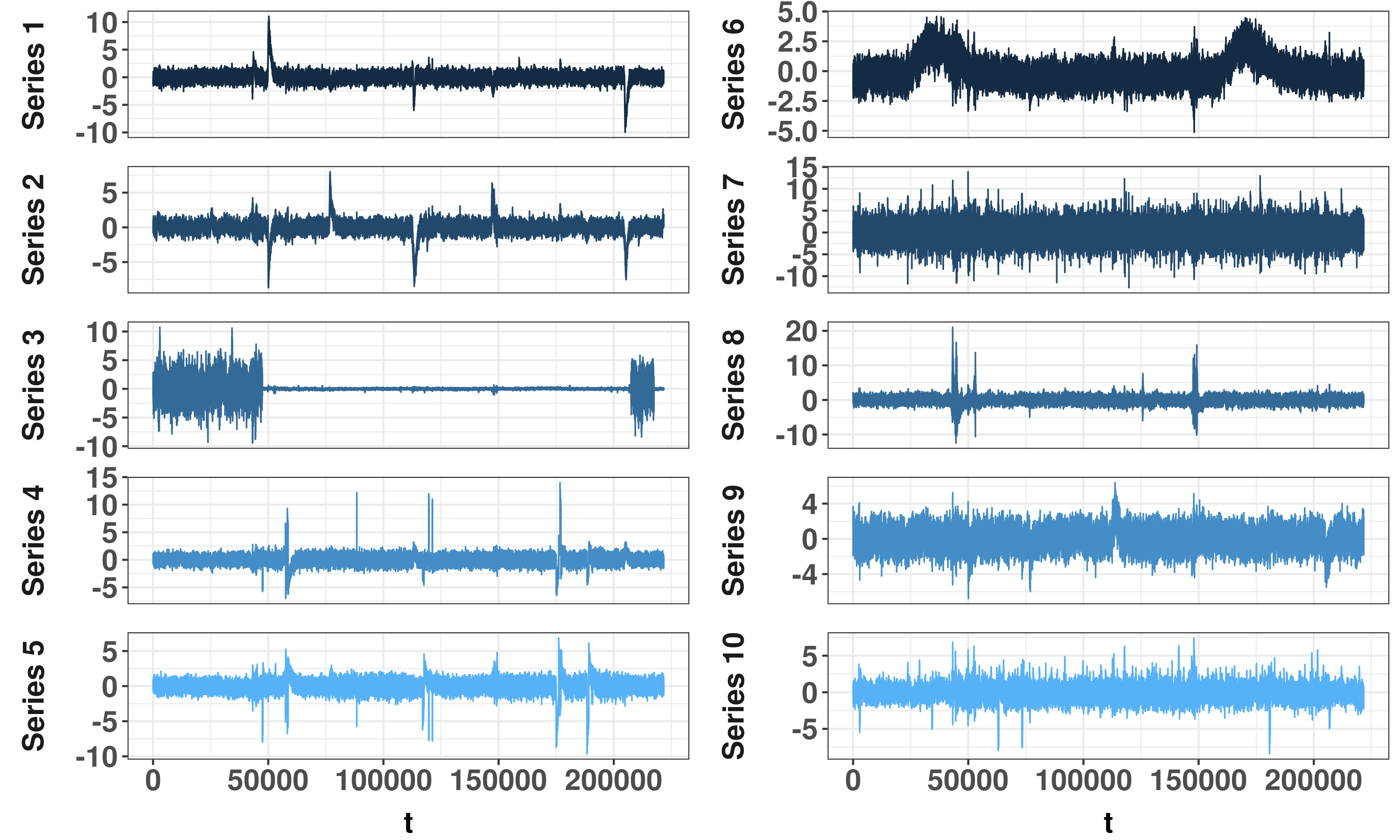}\\
  \caption{Ten nonstationary components recovered by SSAcomb method with $K=12$.}\label{MEGsources}
\end{figure}

Figure~\ref{MEGsources} plots the ten nonstationary components recovered by SSAcomb. 

Due to large number of observations, nonstationarities in autocorrelation are difficult to observe, but nonstationarities in mean and variance are clearly visible in Figure~\ref{MEGsources}. In Figure~\ref{VisA3} in~\ref{App:2} we show the topography plots related to each signal. The topography plots demonstrate which part of the brain is activated and help subject scientists to detect those components that are related to the artifacts and those that are related to the brain activity that is relevant to the study at hand. Notice that many MEG studies proceed by removing artifacts before trying to detect the brain activity signals of interest, see for example~\cite{Escuderoetal:2007} and references therein.     

\section{Conclusion}
\label{sec:discussion}

In this paper we gave a statistical perspective on stationary subspace analysis. We showed how SSA can be formulated as a supervised dimension reduction method, and in that spirit, proposed methods that can detect nonstationarities in mean, variance, autocorrelation, and in all of them. The method thus extended the classical analytic SSA into a more general time series setting. Simulation studies illustrated that the method which combined all three nonstationarity measures outperformed its competitors in various simulation settings. The combined method seemed to perform best when the sample sizes were large and the number of intervals was moderate.  Notice however that performances of different methods depend heavily on the underlying time series components and therefore it might be also of interest to weight the matrices in the combination method if some special structures are of more importance than the others.

In this study we assumed that the number of nonstationary components, $k$, is known. However, most often this parameter value is unknown and needs to be estimated. In \cite{BunauMeineckeKiralyMuller:2009,BlytheBunauMeineckeMuller:2012} a sequential likelihood ratio test was proposed for determining the dimension of the stationary subspace in iid case. In the context of ASSA,~\cite{Haraetal:2010} mentioned that eigenvalues can give guidance for choosing the dimension, but they did not pursue this idea any further. When using SSAsir, SSAsave, SSAcor or SSAcomb, a possible test for testing the null hypothesis $H_0:\ k=k_0$, where $k$ is the true dimension of the nonstationary subspace and $k_0$ is the proposed dimension, could also be based on the eigenvalues of matrices $\bo M$ and resampling-based procedures in a similar fashion as in \cite{NordhausenOjaTylerVirta2017} for example. An estimate for the nonstationary subspace dimension could then be based on the sequential testing as in \cite{VirtaNordhausen2021}. We leave this for a future work. In this context we will also investigate if the detection of the type of nonstationarity can be formalized and how the SSA methods can help to detect change points in the multivariate time series.

\section*{Acknowledgements}\label{sec:acknow}
The work of KN was supported by the Austrian Science Fund (FWF) under grant P31881-N32. We thank Professor Jarmo H\"am\"al\"ainen for providing us the MEG data.


\clearpage

\section*{Appendix}

\appendix

\section{Additional figures comparing SSAsir, SSAsave and SSAcor}
\label{App:1}

Visualizations of SSAsir, SSAsave and SSAcor for the same series as in Section~\ref{sec:CompEx} when the intervals are of unequal lengths.

\begin{figure}[h!]
  \center
  \includegraphics[width=0.8\textwidth]{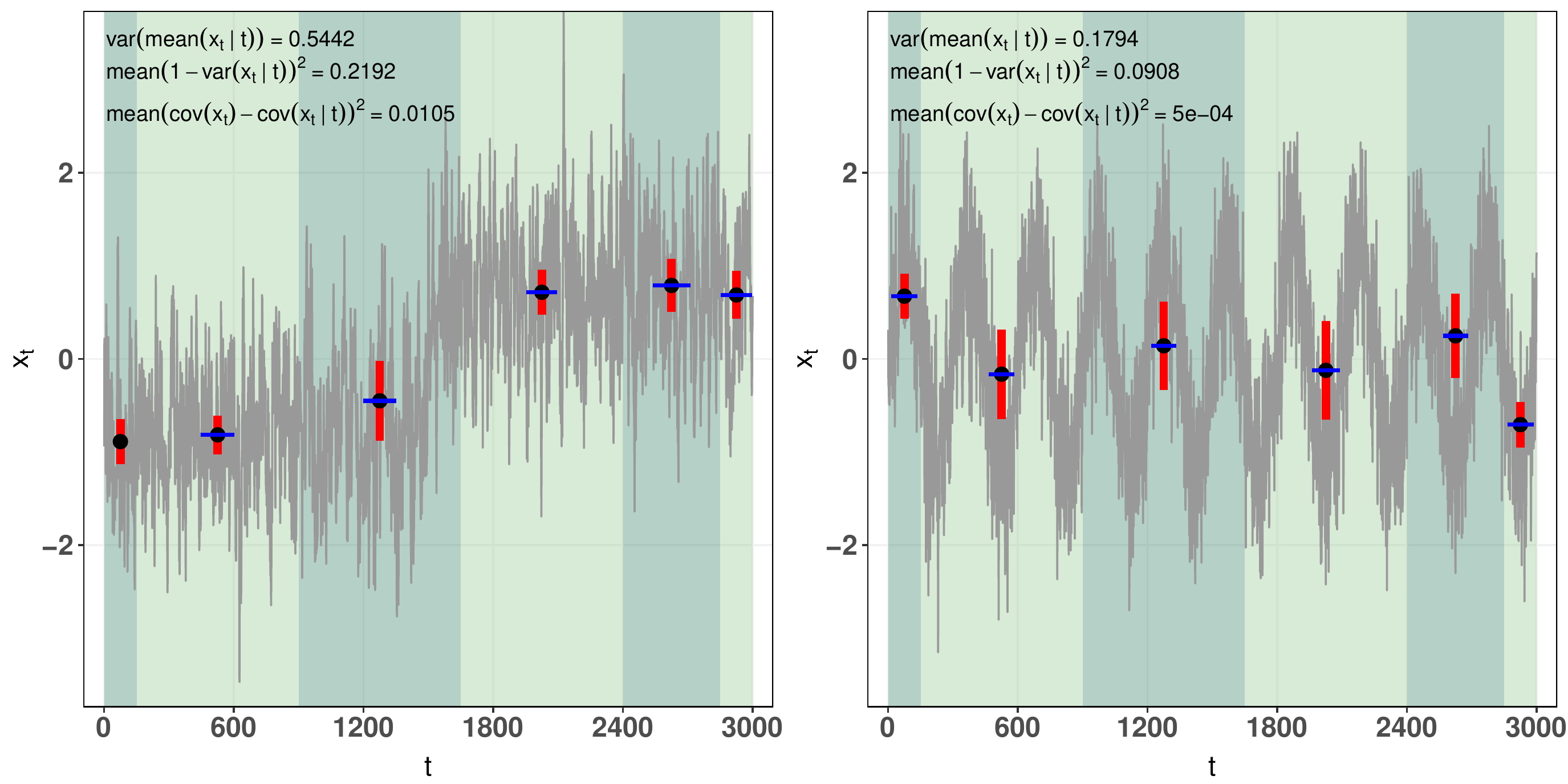}\\
  \caption{Visualization of SSAsir, SSAsave and SSAcor for a component with nonstationarity in mean (left panel trend, right panel seasonality) when the six intervals have different sizes. The black dots represent the interval mean, the height of vertical red bar the interval variance and the width of blue horizontal bar the interval autocovariance.}\label{VisA1}
\end{figure}

\begin{figure}[h!]
  \center
  \includegraphics[width=0.8\textwidth]{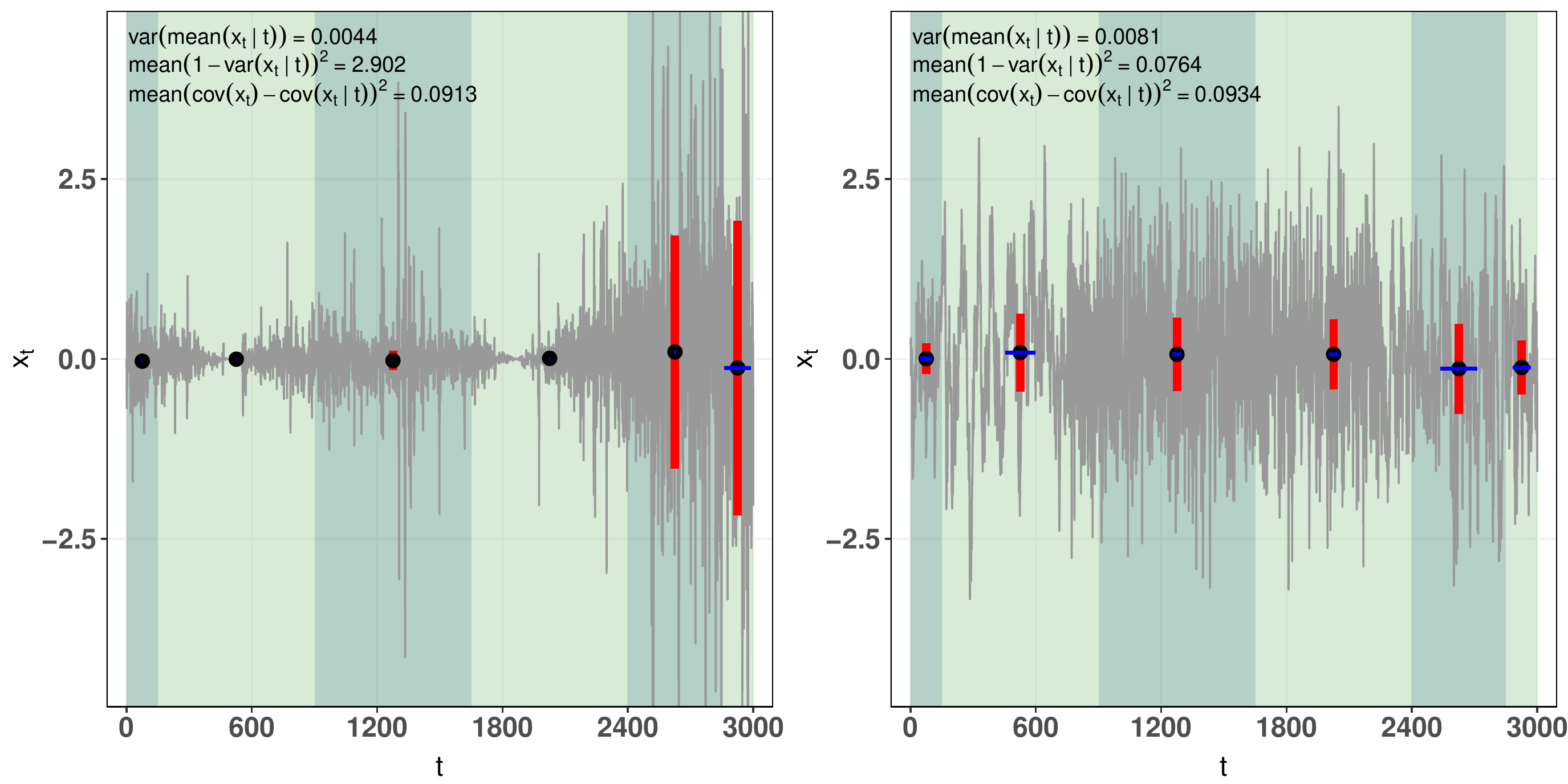}\\
  \caption{Visualization of SSAsir, SSAsave and SSAcor for a component with nonstationarity in variance (left panel) and in autocorrelation (right panel) when the six intervals have different sizes. The black dots represent the interval mean, the height of vertical red bar the interval variance and the width of blue horizontal bar the interval autocovariance.}\label{VisA2}
\end{figure}

\newpage

\section{Additional figures for the MEG example.}
\label{App:2}

\begin{figure}[h!]
  \center
  \includegraphics[width=0.8\textwidth]{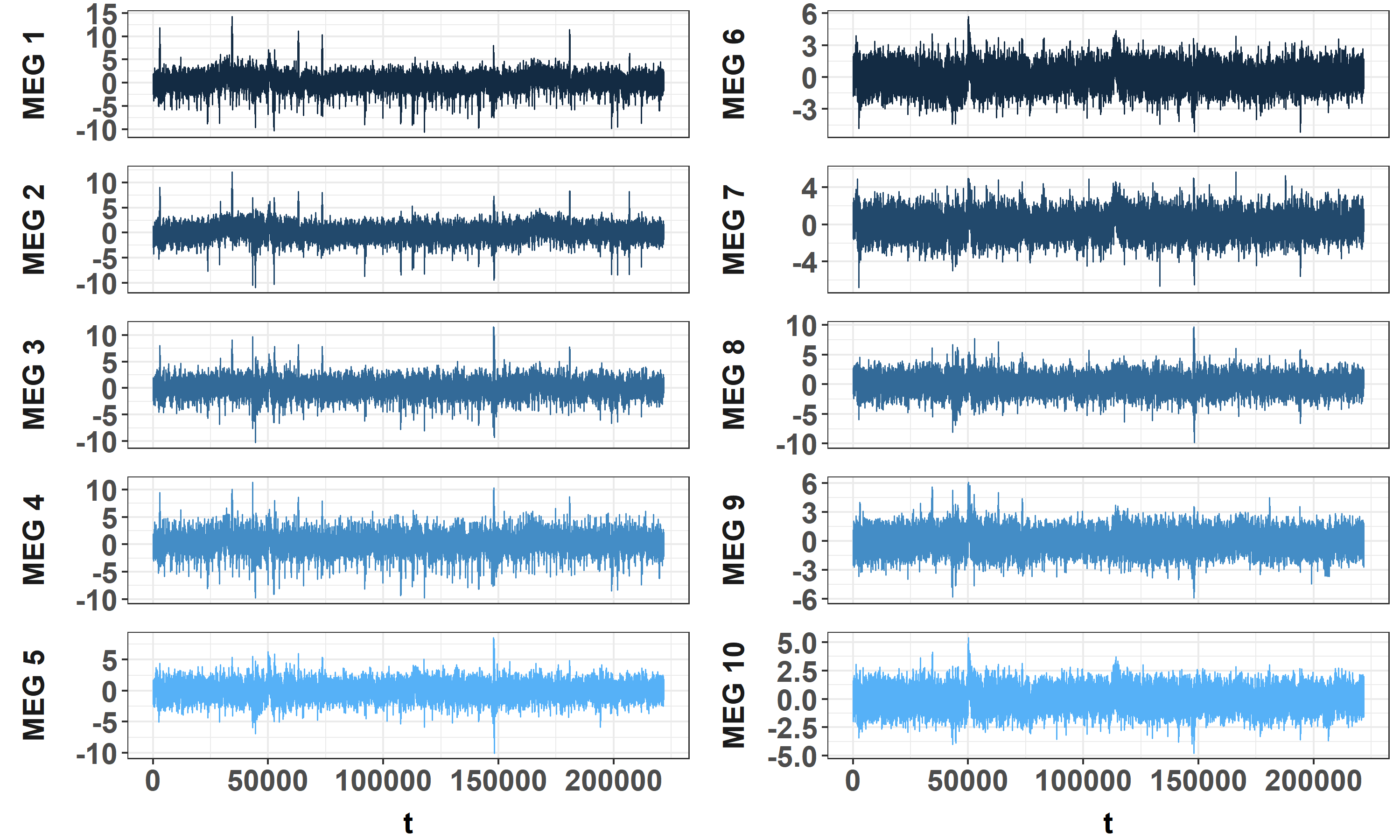}\\
  \caption{Ten observed MEG signals.}\label{MEGoriginal}
\end{figure}

\begin{figure}[ht]
  \center
  \includegraphics[width=0.8\textwidth]{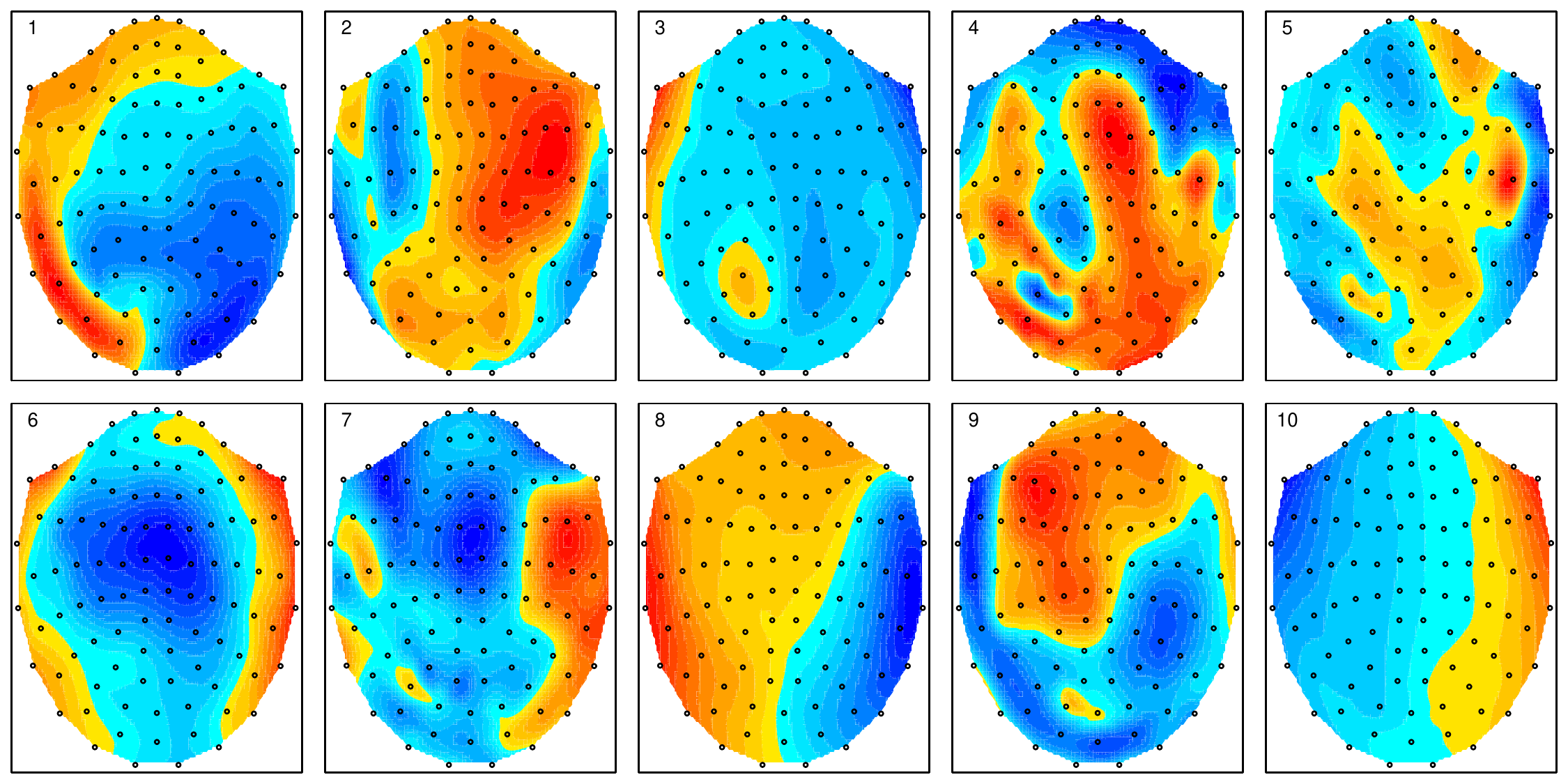}\\
  \caption{Topography plots based on ten nonstationary components recovered from the MEG data. The topographies illustrate which part of the brain is activated related to each of the nonstationary components.}\label{VisA3}
\end{figure}



\end{document}